# BUCKLING-INDUCED TRANSMISSION SWITCHING IN PHONONIC WAVEGUIDES IN THE PRESENCE OF DISORDER

Ali Kanj[1], Alexander F. Vakakis[1], Sameh Tawfick[1,2]

[1]Department of Mechanical Science and Engineering, University of Illinois at Urbana-Champaign, Illinois 61801, United States

[2]The Beckman Institute of Advanced Science and Technology, University of Illinois at Urbana-Champaign, Illinois 61801, United States

**On-chip phononic circuits tailor the transmission of elastic waves, which can couple to electronic and photonic systems, enabling new signal manipulation capabilities. Phononic circuits rely on waveguides that transmit elastic waves within desired frequency passbands, typically designed based on the Bloch modes of the waveguide constitutive cell, assuming linearity and periodicity. MEMS waveguides composed of coupled drumhead (membrane) resonators offer tunable MHz operation frequencies for applications in nonlinear optomechanics, topological insulators, phononic cavities, and acoustic switching. Here, we construct a reduced-order model (ROM) to demonstrate the switching of signal transmission in drumhead-resonator waveguides due to thermoelastic buckling. The ROM shows that buckling amplifies existing structural disorders, breaking the periodicity required for waveguide transmission through the first passband. This periodicity breaking manifests in the localization of the first-passband modes, like classical Anderson localization caused by disorders. The proposed ROM is essential to study the investigated phenomena since Bloch mode analysis fails for weakly-disordered (< 5%) finite waveguides due to the disorder amplification caused by the thermoelastic buckling. The illustrated transmission control should be useful for logical acoustic operations, like switching, and can be extended to 2D circuits in the future.**

## I. Introduction

Phononic circuits are gaining increased interest because they tailor the propagation of elastic and acoustic waves, which is advantageous for signal manipulation. For example, phononic circuits are useful for cellular phone duplexers by serving as acoustic isolators and mirrors [1] [2] [3]. In medical ultrasound applications and acoustic nondestructive tests, phononic circuits promise to miniaturize the imaging aperture [4] [5] [6], decouple the electro-acoustic transduction [7] [8], and slow the signal for smaller delay lines [9] [10] [11]. Moreover, nanostructural phononics operating in the hypersonic (GHz to THz) frequencies enable thermal management [12] [13] [14], photonic-phononic interactions [15] [16], and quantum information control [17] [18]. Phononic structures offer readily-achievable nonlinearities allowing for strong optomechanical nonlinearities [19] [20], targeted-energy transfer [21] [22], and passive structural nonreciprocity [23] [24] [25].

Phononic circuits require accurately designed and fabricated waveguides to spatially constrain the acoustic transmission within a specific frequency range referred to as the passband (or the transmission band). In the passband, the temporal frequencies are linked to the spatial frequencies (i.e., the wavenumbers) through the dispersion relation of the medium, providing additional control over the acoustic transmission [1] [26]. This temporal and spatial selectiveness stems from the dynamic characteristics of the unit cells whose periodic repetition forms the waveguide. Therefore, the unit cell design is directly linked to the waveguide characteristics via the Bloch modes of the unit cell. The Bloch modes are the vibrational modes that the unit cell exhibits under Floquet boundary conditions with a wavenumber spanning the irreducible Brillouin zone (IBZ) [26]. This approach calculates the possible wavenumber-frequency relationship known as the band structure of the phononic crystal (i.e., the unit cell). This band structure matches the transmission in an *infinite periodic* waveguide of the *same repeated* unit cell [26].

Bloch modes predict the transmission of sufficiently long and weakly-disordered waveguides [5] [6] [12] [16] [27] [28], although fabricated waveguides are neither infinite nor perfectly periodic. In these cases, the finite-structure modal frequencies lie within (or close to) the Bloch modes passbands [26]. For example, such a waveguide of $N$ cells possesses at most $N$ finite-structure modes for every passband; increasing $N$ makes the $N$ modes more densely packed within the passband leading to the continuous Bloch-modes band structure as $N \to +\infty$. The dense packing of modes originates from the structural periodicity whose absence (i.e., aperiodicity) generates frequency-distinct modes that cannot approximate the passbands. In addition, the periodicity causes (spatially-) extended mode shapes that permit the transmission of a signal between the ends of the structure [26]. These features – the approximate passband and the extended mode shapes – are acoustically attractive and enable a finite periodic structure to operate as a waveguide. The Bloch mode approach is computationally efficient in linear periodic systems because it enables the tailoring of a single unit cell to estimate the behavior of the entire waveguide. On the other hand, it significantly deviates from experimental results when the number of unit cells is limited, when there is aperiodicity (structural asymmetry) in the devices (whether intentional or uncontrolled), and when nonlinearities are profound.

Considering repetitive arrays of drumhead resonators composed of coupled flexible micro-membranes, we have recently shown that thermoelastic buckling of the membranes can switch the acoustic transmission [29]. Waveguides made from coupled drumhead resonators were first proposed by Hatanaka et al. in 2013 [30], who showed that they sustain megahertz-to-gigahertz mechanical vibrations with high quality factors (high $Q$s) and optical finesse, features which are valuable in mechanical, electrical, and optical applications [31, 32, 33]. For instance, optomechanical interactions favor large surface-area structures (like the drumhead resonators) over beams/cantilevers [32, 33, 34]. Another advantage of the drumhead resonators is their manufacturability via conventional micro/nanofabrication [32, 33], while allowing for *in-situ* structural tunability and actuation via piezoelectric [30, 34], electrostatic [35], and thermal control [29]. Therefore, drumhead resonators were applied in tunable optical cavities [36] and low-loss nonlinear optomechanical coupling [37]. Moreover, coupling drumhead resonators in the form of arrays, like the devices studied in this article, served in realizing phononic transistors [30], tunable 1D phononic waveguides [35], cavity-switchable waveguides [38], and on-chip 2D topological insulators [39].

In this work, we study the mechanism of transmission switching in the drumhead-resonator waveguides reported in [29], a phenomenon that has previously been attributed to buckling-induced aperiodicity. Specifically, we develop a reduced-order model (ROM) that mimics the experiments observed in [29] (section II). The ROM accounts for out-of-plane translation, rotation, and coupling to accurately predict the first and second passbands of the waveguides as functions of the buckling state. The ROM uses the concept of the *von* Mises truss [40] to capture the effect of buckling on drumhead-resonator waveguides, as illustrated in the electro-thermoelastic tunability of individual drumhead resonators in [41]. In turn, the *von* Mises trusses permit modeling and predictive analysis of the drumhead-resonator waveguides via lumped springs and rigid bodies, presenting simpler models that are amenable to analytical studies compared to finite element models (e.g., continuous beams on elastic foundations).

With the *von* Mises ROM, we calculate the Bloch modes (section III), and compare them to the transmission of (60-cell) *finite* waveguides in cases of *perfect periodicity* and (< 5%) *weak disorder* (section IV). We investigate the acoustics of the finite waveguides by subjecting their first cell to nonzero initial velocities and monitoring the resulting free responses in the time and frequency domains as functions of the spatial propagation of wave packets in the waveguide. We find that when the *weakly-disordered* finite waveguides are close to their critical buckling state, the transmission through the first passband vanishes. Stronger disorder results in a larger range of temperatures where the first passband does not transmit elastic waves. This contrasts with the corresponding *perfectly-periodic* finite waveguide (i.e., with no disorder), where the first passband transmits elastic waves at all considered temperatures, even at the onsite of critical buckling. As for the second passband, the acoustic transmission persists for all considered disorders and temperatures.

To thoroughly explain the effect of buckling on the transmission, we inspect the dependencies of the mode shapes of the considered waveguides on temperature (section V). The results show that the transmission-switching is associated with converting the mode shapes from extended over the entire waveguide to localized at some cells. This localization of mode shapes with disorders conforms to Anderson's localization originally discovered in electromagnetic waves [42] [43] [44] and then applied in elastic settings [27] [28] [45]. Finally, we present an evaluation of this buckling-switchable transmission on a finite element model (FEM) of the experimental waveguide studied in [29] with 5% disorder far from, or close to critical buckling (supplemental Video.S2). The FEM simulations agree with the predictions of the ROM, thus conclusively proving that weak disorder leads to loss of transmission in the repetitive array of drumhead resonators due to buckling.

## II. Description of the waveguide and the reduced-order model (ROM)

In this work, we study the phononic waveguides shown in Fig. 1a. This waveguide consists of repetitive cells capable of transmitting flexural acoustic waves [29] [38] [46] [47]. This waveguide was studied in [29], where the cells are drumhead-like membranes composed of Silicone Nitride ($SiN_x$) suspended by an etched Silicone Oxide ($SiO_2$) layer on top of a Silicone (Si) substrate. The involved materials and fabrication methods induce residual stresses in the waveguide, whose cells buckle as depicted in Fig. 1b by atomic force microscopy (AFM) conducted in [29].

In Fig. 1c, we show the effect of buckling on the elastic transmission of the waveguide of Fig. 1a [29]. The temperature in Fig. 1c controls the state of buckling in the waveguide, where lower temperature increases compressions between cells to provoke stronger buckling. At each temperature, the colormap in Fig. 1c corresponds to the frequency response measured at the middle cell of the waveguide due to the electrostatic actuation of the gold (Au) pad covering the first cell (cf. Fig. 1a). At high temperatures in Fig. 1c (i.e., above ~230 K), the waveguide exhibits three frequency regimes of effective transmission corresponding to the first three passbands (labeled as I, II, and III). A decrease in temperature from 280 K down to ~230 K decreases the mean frequency of all the passbands, indicating a *softening* behavior. During this softening, passband I diminishes its bandwidth until collapsing at ~230 K, whereas passbands II and III maintain an almost constant bandwidth. *Reducing the temperature to below ~230 K completely eliminates passband I* and increases the mean frequencies of passbands II and III while possessing almost constant bandwidths. The observed temperature-dependent changes in the frequency passbands and the switch in frequency detuning imply that the waveguide at ~230 K is in a critical buckling state associated with the softest structural configuration (since buckling indicates minimum linearized stiffness). Accordingly, the waveguide is pre-buckled for temperatures > ~230 K and post-buckled for temperatures < ~230 K, based on [29]. In both buckling regimes, the frequency detuning of passbands II and III are direct consequences of the buckling state of the waveguide. *However, the frequency detuning* of *passband I and its* *transmission loss in the post-buckled regime necessitates both buckling and disorder in the waveguide* [29].

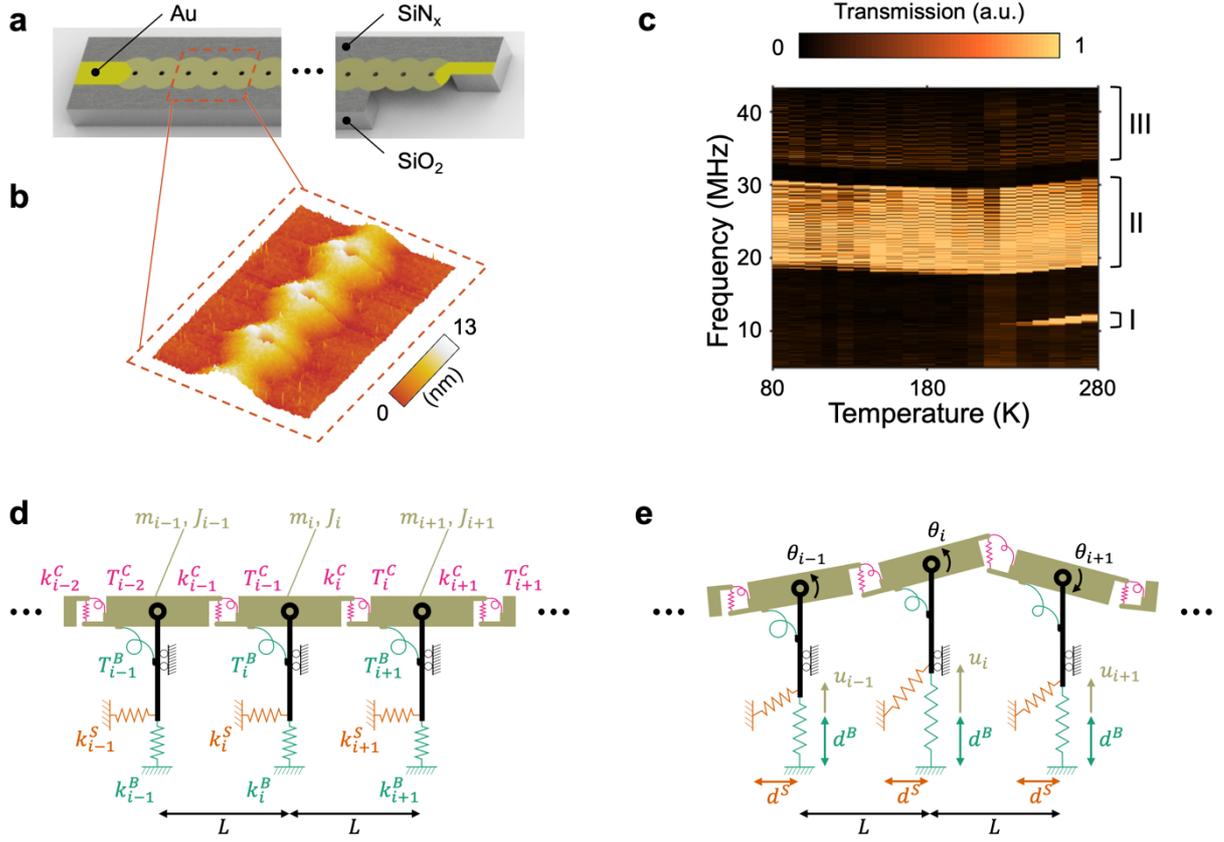

Fig. 1. Thermally buckled elastic waveguide: (a) Schematic drawing of a MEMS phononic waveguide made of coupled drumhead-resonators [29]; (b) 3D topography map of 3 cells of the waveguide [29] measured using atomic force microscopy (AFM) – the colormap shows the out-of-plane deflections resulting from buckling in the structure; (c) measured transmission in the waveguide [29] as a function of temperature and frequency of excitation applied to the first cell in the waveguide – the colormap depicts the amplitude of oscillations at the middle of the waveguide, which shows that the temperature change eliminates the transmission in passband I and detunes the frequency of passbands I, II, and III; schematic drawings of the proposed reduced-order model (ROM) of the waveguide undergoing thermal buckling showing (d) undeformed and (e) deformed states.

To further investigate this relationship between disorder, buckling, and elastic transmission in the considered waveguide, we propose the reduced-order model (ROM) depicted in Figs. 1d-e. This ROM captures the thermally-mediated elastic buckling based on the ROM of a single cell introduced in [41], exhibiting very good predictive capacity. Here, we extend the ROM of [41] to account for the coupling between the cells in the waveguide and model the acoustics of the entire phononic waveguide. Accordingly, we allocate to each cell a translational degree-of-freedom (DoF) (as in [41]) and a rotational DoF to capture passbands I and II, respectively.

As shown in Fig. 1d, each cell of index $i$ in the waveguide consists of a rigid mass $m_i$ with a moment of inertia $J_i$. Cell $i$ undergoes the motion illustrated in Fig. 1e with translational coordinate $u_i$ and rotation angle $\theta_i$. The translation deforms the grounding springs of stiffnesses $k_i^B$ and $k_i^S$ representing the restoring forces for bending and stretching, respectively. As in [41], these translational bending and stretching springs are confined at distances $d^B$ and $d^S$ (see Fig. 1e) while possessing free (undeformed) lengths $L^B$ and $L^S$, respectively; clearly, a free length larger than the confinement distance (i.e., $L^B > d^B$ and $L^S > d^S$) introduces compressive strains (precompression) in the cell. We assume that the remaining springs in the ROM are undeformed at their undeformed positions. For example, the springs with stiffnesses $T_i^B$, $k_{i-1}^C$, $k_i^C$, $T_{i-1}^C$, and $T_i^C$ attached to the cell of index $i$ don't apply any forces or torques in Fig. 1d. The grounding torsional spring of stiffness $T_i^B$ lumps the bending effects that oppose the rotation $\theta_i$ due to the grounded boundary of the drumhead. The coupling springs with stiffnesses $k_i^C$ and $T_i^C$ account for the force and torque, respectively, applied by cell $i$ to cell ($i + 1$) due to the deformations illustrated in Fig. 1e. Lastly, we represent the lattice length separating two successive cells by the length $L$ (see Figs. 1d-e).

### III. Bloch modes of a single cell

In a previous article [41], we described the static equilibrium and the equations of motion of a single drumhead resonator and identified its system parameters, which we refer to as the reference cell parameters. These parameters are the translating mass $m_{Ref}$ and springs $k_{Ref}^B$ and $k_{Ref}^S$ (we use the subscript "$Ref$" to label the reference cell), which are reproduced in Table I. We start by studying the Bloch modes of an infinite waveguide based on a repetition of this reference unit cell, as shown in Fig. 1d-e. The grounding translating springs exert the force $F_i^{Buck}$ at the cell $i \in \{1, 2, ...\}$ and the temperature $T$ expressed for any translational displacement $u_i$ as:

$$F_i^{Buck}(u_i; T) = k_i^B d^B \left[\frac{u_i}{d^B} - \delta^B(T)\right] + k_i^T u_i \left[1 - \frac{1 + \delta^S(T)}{\sqrt{1 + \left(\frac{u_i}{d^S}\right)^2}}\right]. \quad (1)$$

In (1), $\delta^B(T)$ and $\delta^S(T)$ are the temperature-dependent bending and stretching strains, respectively, stemming from the thermal expansion and the fabrication-residual stresses in the cells. We characterize these strains by the following temperature dependencies (as discussed in [41]),

$$\delta^B(T) \stackrel{\text{def}}{=} \frac{d^B - L^B}{L^B} = \beta_0 + \beta_1 T + \beta_2 T^2 \quad (2a)$$

$$\delta^S(T) \stackrel{\text{def}}{=} \frac{d^S - L^S}{L^S} = \gamma_0 + \gamma_1 T, \quad (2b)$$

with the values of $\beta_0, \beta_1, \beta_2, \gamma_0$, and $\gamma_1$ listed in Table I. We assume that these temperature dependencies govern the strains of all the cells in the waveguide. Moreover, we nondimensionalize the forces in this work by $k_{Ref}^B d^B$ leading to the following nondimensional buckling force,

$$\bar{F}_i^{Buck}(\bar{u}_i; T) = \bar{u}_i - \delta^B(T) + \kappa_i^T \bar{u}_i \left[1 - \frac{1 + \delta^S(T)}{\sqrt{1 + \left(\frac{\bar{u}_i}{\bar{d}^S}\right)^2}}\right], \quad (3)$$

where $\kappa_i^T \stackrel{\text{def}}{=} k_i^T/k_i^B$, $\bar{d}^S \stackrel{\text{def}}{=} d^S/d^B$, and $\bar{u}_i \stackrel{\text{def}}{=} u_i/d^B$ with the overbar denoting a nondimensionalized entity.

Focusing on the reference cell which undergoes only translation while connected to $k_{Ref}^B$ and $k_{Ref}^S$, we find its equilibrium displacement $\bar{u}_{Ref}^{Eqm}(T)$ by solving the following equation:

Table I. Parameters of the reference cell [41].

| $\delta^B$ | | | $\delta^S$ | | $\bar{d}^S$ | $\kappa_{Ref}^T$ | $\frac{1}{2\pi}\sqrt{\Lambda_{Ref}^B}$ [MHz] | $\chi$ |
|---|---|---|---|---|---|---|---|---|
| $\beta_0$ | $\beta_1$ [K$^{-1}$] | $\beta_2$ [K$^{-2}$] | $\gamma_0$ | $\gamma_1$ [K$^{-1}$] | | | | |
| 7.65 | -3.47E-2 | 3.81E-5 | 1.9 | -4.07E-3 | 1 | 1 | 9.40 | 1/12 |

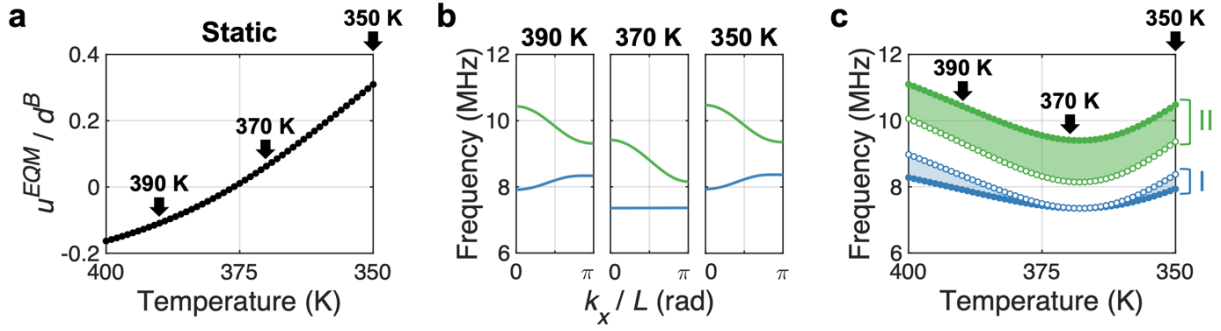

Fig. 2. Thermal buckling of the *infinite perfectly periodic* waveguide (i.e., Bloch modes): (a) Static equilibrium of a single cell as a function of temperature based on (4); (b) frequency dispersion curves as a function of the nondimensional wavenumber of the Bloch modes at a temperature of 390 K, 370 K, and 350 K; (c) Bloch-modes frequency extrema as a function of temperature illustrating the transmission detuning in an infinite perfectly-periodic waveguide – we depict the Bloch-modes frequency extrema with $k_x/L$ close to 0 rad by the filled circles, whereas the extrema with $k_x/L$ close to $\pi$ rad by open circles; also the blue and green colors in (b, c) represent the Bloch-mode passbands I and II, respectively.

$$\bar{F}_{Ref}^{Buck}\left(\bar{u}_{Ref}^{Eqm}; T\right) = 0 \text{ with the maximum satisfying } \left.\frac{d\bar{F}_{Ref}^{Buck}}{d\bar{u}_{Ref}}\left(\bar{u}_{Ref}; T\right)\right|_{\bar{u}_{Ref}=\bar{u}_{Ref}^{Eqm}} > 0 \quad (4)$$

The maximum condition in (4) ensures $\bar{u}_{Ref}^{Eqm}$ to be the most stable equilibrium solution, which should be favored experimentally. In Fig. 2a, we plot the values of $\bar{u}_{Ref}^{Eqm}$ as a function of temperature based on (4), (3), (2), and the reference parameters in Table I. We observe that the single cell translates upwards due to cooling, which increases the internal compressions leading to buckling of the cell [41].

To study the effect of buckling on wave transmission, we evaluate the Bloch modes of the cell at each temperature shown in Fig. 2a. The Bloch modes correspond to the infinite waveguide of Fig. 1e-d made of cells whose parameters are *identical* to the considered single cell. In this *perfectly periodic infinite* waveguide, all the cells attain at $T$ the equilibrium state of $\bar{u}_i^{Eqm} = \bar{u}_{Ref}^{Eqm}(T)$ and $\theta_i^{Eqm} = 0$ rad for all $i \in \{1,2,3,\ldots,+\infty\}$. At every instant $t$, we track the oscillations of the $i^{th}$ cell about its equilibrium state via the perturbation coordinates:

$$\bar{v}_i(t) \stackrel{\text{def}}{=} \bar{u}_i(t) - \bar{u}_i^{Eqm}, \quad (5a)$$

$$\bar{h}_i(t) \stackrel{\text{def}}{=} \bar{L}\theta_i(t) - \bar{L}\theta_i^{Eqm}, \text{ where } \bar{L} \stackrel{\text{def}}{=} L/d^B. \quad (5b)$$

The above coordinates allow writing Newton's second law on any cell of index $i > 1$ in Fig. 1d-e as,

$$\mu_i \begin{bmatrix} 1 & 0 \\ 0 & \chi \end{bmatrix} \begin{Bmatrix} \frac{d^2 \bar{v}_i}{dt^2} \\ \frac{d^2 \bar{h}_i}{dt^2} \end{Bmatrix} + \mu_{i-1} \begin{bmatrix} -\Lambda_{i-1}^C & -\frac{\Lambda_{i-1}^C}{2} \\ \frac{\Lambda_{i-1}^C}{2} & \frac{\Lambda_{i-1}^C}{4} - \Gamma_{i-1}^C \end{bmatrix} \begin{Bmatrix} \bar{v}_{i-1} \\ \bar{h}_{i-1} \end{Bmatrix} +$$

$$\left( \mu_{i-1} \begin{bmatrix} \Lambda_{i-1}^C & -\frac{\Lambda_{i-1}^C}{2} \\ -\frac{\Lambda_{i-1}^C}{2} & \frac{\Lambda_{i-1}^C}{4} + \Gamma_{i-1}^C \end{bmatrix} + \mu_i \begin{bmatrix} \Lambda_i^C + \Lambda_i^{Buck} & \frac{\Lambda_i^C}{2} \\ \frac{\Lambda_i^C}{2} & \frac{\Lambda_i^C}{4} + \Gamma_i^C + \Gamma_i^B \end{bmatrix} \right) \begin{Bmatrix} \bar{v}_i \\ \bar{h}_i \end{Bmatrix} + \quad (6)$$

$$\mu_i \begin{bmatrix} -\Lambda_i^C & \frac{\Lambda_i^C}{2} \\ -\frac{\Lambda_i^C}{2} & \frac{\Lambda_i^C}{4} - \Gamma_i^C \end{bmatrix} \begin{Bmatrix} \bar{v}_{i+1} \\ \bar{h}_{i+1} \end{Bmatrix} = \begin{Bmatrix} 0 \\ 0 \end{Bmatrix},$$

where $\mu_i \stackrel{\text{def}}{=} m_i/m_{Ref}$, $\chi \stackrel{\text{def}}{=} J_i/m_i L^2$, $\Lambda_i^{Buck}(T) \stackrel{\text{def}}{=} \Lambda_i^B \frac{d\bar{F}_i^{Buck}}{d\bar{u}_i}\Big|_{\bar{u}_i = \bar{u}_i^{Eqm}(T)}$, $\Lambda_i^B \stackrel{\text{def}}{=} k_i^B/m_i$, $\Lambda_i^C \stackrel{\text{def}}{=} k_i^C/m_i$, $\Gamma_i^C \stackrel{\text{def}}{=} T_i^C/(m_i L^2)$, and $\Gamma_i^B \stackrel{\text{def}}{=} T_i^B/(m_I L^2)$. Equation (6) only considers the linearized dynamics of the undamped $i^{\text{th}}$ cell. Note that the nondimensionalization in (6) results in (squared) frequency-like parameters (i.e., $\Lambda_i^{Buck}$, $\Lambda_i^C$, $\Gamma_i^B$, and $\Gamma_i^C$). This parameter conversion offers an advantage when comparing the model to experiments because frequencies are easier to identify than stiffnesses and directly affect the performance of the waveguides. For instance, we deduce the value of $\Lambda_{Ref}^B$ listed in Table I from the experiments of a single cell in [41]. For the remaining frequency-like parameters, we assume the following relationships for all $i \in \{Ref, 1, 2, 3, \dots\}$:

$$\Lambda_i^C(T) = 0.2 \left[ \Lambda_i^G(T) - \min_{350 \to 400 \text{ K}} \Lambda_i^G(T) \right] \quad (7a)$$

$$\Gamma_i^B = \frac{1}{12} \Lambda_i^B \quad (7b)$$

$$\Gamma_i^C(T) = \frac{1}{12} \left[ 3\Lambda_i^C(T) - \frac{3}{4} \Gamma_i^B \right]. \quad (7c)$$

To calculate the Bloch modes of a single cell, we apply the Floquet boundary conditions of $\begin{Bmatrix} \bar{v}_i \\ \bar{h}_i \end{Bmatrix} = \boldsymbol{p}_i e^{j\left(\frac{k_x}{L}i + \omega t\right)}$ with a normalized wavenumber $k_x/L$, a modal frequency of $\omega$, a modal vector $\boldsymbol{p}_i$, and the imaginary number $j^2 = -1$. Additionally, we assume that all cells are ***identical*** to the single cell with parameters listed in Table I, transforming (6) into the following boundary value problem:

$$\left( -\omega^2 \begin{bmatrix} 1 & 0 \\ 0 & \chi \end{bmatrix} + \begin{bmatrix} 2\left(1 - \cos\frac{k_x}{L}\right) \Lambda_{Ref}^C + \Lambda_{Ref}^{Buck} & \cdots \\ -\Lambda_{Ref}^C \sin\frac{k_x}{L} & \end{bmatrix} \right. \quad (8)$$

$$\cdots \frac{1}{2}\left(1+\cos\frac{k_x}{L}\right)\Lambda^C_{Ref} + 2\left(1-\cos\frac{k_x}{L}\right)\Gamma^C_{Ref} + \Gamma^B_{Ref}\right]\right)\boldsymbol{p}_i = \begin{Bmatrix}0\\0\end{Bmatrix}.$$

$$\Lambda^C_{Ref}\sin\frac{k_x}{L}$$

For all $k_x/L \in [0, \pi]$ rad, the irreducible Brillouin zone (IBZ) is defined by the respective pair of eigenfrequencies $\omega$ that zero the determinant of the matrix operating on $\boldsymbol{p}_i$ in (8). These eigenfrequencies are the Bloch modes' frequencies forming the dispersion curves in Fig. 2b at 390 K, 370 K, and 350 K for the single cell. The lower (blue) and upper (green) curves in Fig. 2b correspond to passbands I and II of the transmission in the *perfectly periodic infinite* waveguide, respectively. We depict the transmission of this waveguide in Fig. 2c, where we collect the extrema (maxima and minima) of passbands I and II (like in Fig. 2b) for the temperature $T \in [350, 400]$ K.

Fig. 2c shows that cooling from 400 to ~370 K reduces the mean frequencies of both passbands while narrowing the bandwidth of passband I. Cooling below ~370 to 350 K increases again the mean frequencies of both passbands while widening the bandwidth of passband I. The first cooling phase from 400 to ~370 K in Fig. 2c resembles the cooling phase in Fig 1c between 280 and ~230 K. However, the second cooling phase between ~370 to 350 K in Fig. 2c diverges fundamentally from the experimental transmission in Fig. 1c between ~230 and ~80 K, where the transmission in passband I does not reemerge. Therefore, the ROM buckling cannot eliminate the transmission of passband I in a *perfectly periodic infinite* waveguide. This loss of transmission with buckling necessitates the consideration of disorder (i.e., the break of perfect periodicity) in the waveguide as previously established in [29].

Note that we adopt the relationships in (7) to emulate the experimental transmission in Fig 1c between 280 and ~230 K. For this reason, we select $\Lambda^C_i(T)$ in (7a) to decrease until the transmission vanishes at the point of minimum frequency, $\min_{350\to 400\text{ K}} \Lambda^C_i(T)$, leading to the shrinkage of passband I between 400 and ~370 K in Fig. 2c. In (7b), we assume that the rotation of the cell centerline (of length $L$) deflects an elastic foundation of stiffness density $k^B_i/L$. In (7c), we impose a temperature-constant bandwidth for passband II like the measurements in Fig. 1c. The temperature-detuning of the mean frequencies of the passbands in Fig. 2c is considered for the identified parameters (i.e, $\Lambda^B_{Ref}$, $\beta_0$, $\beta_1$, $\beta_2$, $\gamma_0$, and $\gamma_1$) of the ROM in [41], which slightly deviate from those in the devices used in Fig. 1c (extracted in [29]).

## IV. Temporal transmission in finite waveguides

We focus now on the waveguide disorder resulting from the thickness variation between the cells. We assume a thickness variation of the form,

$$\frac{h_i - h_{Ref}}{h_{Ref}} = \frac{\sigma_h}{4}\left\{\underbrace{\left(2\frac{\left|\frac{N+1}{2} - i\right|}{\frac{N+1}{2} - 1} - 1\right)}_{s_i} + r_i([-1,1])\right\}, \quad (9)$$

for $i \in \{1,2,\ldots,N\}$, where we denote by $h_i$ the thickness of the $i^{\text{th}}$ cell, $h_{Ref}$ the thickness of the reference cell discussed in the previous section, $\sigma_h$ the level of thickness disorder, and $r_i([-1,1])$ a random rational number $\in [-1,1]$ generated at each $i$. We introduce the random number $r_i$ to account for the random errors of the fabrication process. The $s_i$ term in (9) represents the systematic errors resulting from wet etching that forms the waveguide cells as explained in [29] [38] [46] [47].

The holes at the center of the cells in Fig. 1a-b are etching holes through which the etchant attacks the underlying layer and suspends the cells. Thus, there is a higher (linear) density of etching holes at the middle of the waveguide (of index $\frac{N+1}{2}$) compared to the ends (of indices 1 and $N$). This higher etching-holes density increases the etching rate leading to over-etching at the middle of the waveguide compared to its ends [29]. We model this over-etching by $s_i$ in (9) as a linear distribution of the cell position from the middle of the waveguide. Figs. 3a-b show two examples of thickness variation with $\sigma_h = 0\%$ (i.e., *perfectly periodic* waveguide) and $\sigma_h = 5\%$, respectively.

The thickness variation implies a corresponding variation in the dynamical properties of the cells. Based on the theory of the mechanics circular plates [48] [49] and the assumption in [41], the thickness affects the parameters of the $i^{\text{th}}$ cell in Fig. 1d-e as follows:

$$\frac{\kappa_i^S}{\kappa_{Ref}^S} = \left(\frac{h_i}{h_{Ref}}\right)^{-2}, \quad (10a)$$

$$\frac{\Lambda_i^B}{\Lambda_{Ref}^B} = \left(\frac{h_i}{h_{Ref}}\right)^{2}. \quad (10b)$$

The scaling relationships (10) with the expressions in (7) quantify the effect of the cell's thickness on the ROM parameters.

With the ROM of Fig. 1d-e, we apply Newton's 1$^{\text{st}}$ law to calculate the static equilibrium $(\bar{u}_i^{Eqm}, \bar{L}\,\theta_i^{Eqm})$ of each cell $i \in \{1, 2, \ldots, N\}$ in the $N$ cells waveguide using,

$$\underbrace{\begin{Bmatrix} \bar{F}_1^{Buck}(\bar{u}_1^{Eqm}; T) \\ 0 \\ \vdots \\ \bar{F}_i^{Buck}(\bar{u}_i^{Eqm}; T) \\ 0 \\ \vdots \\ \bar{F}_N^{Buck}(\bar{u}_N^{Eqm}; T) \\ 0 \end{Bmatrix}}_{\bar{Q}^{Buck}} + \bar{K}^{Stat} \underbrace{\begin{Bmatrix} \bar{u}_1^{Eqm} \\ \bar{L}\theta_1^{Eqm} \\ \vdots \\ \bar{u}_i^{Eqm} \\ \bar{L}\theta_i^{Eqm} \\ \vdots \\ \bar{u}_N^{Eqm} \\ \bar{L}\theta_N^{Eqm} \end{Bmatrix}}_{\bar{q}^{Eqm}} = \mathbf{0}, \qquad (11)$$

where $\bar{F}_i^{Buck}(\bar{u}_i^{Eqm}; T)$ is the thermo-elastic buckling force expressed in (3). In (11), we assume small angles of deformation allowing the approximation $\sin \theta_i^{Eqm} \approx \theta_i^{Eqm}$ while neglecting the longitudinal displacement of the cells' ends. Under this approximation, we express the nondimensional static stiffness $\bar{K}^{Stat}$ as,

$$\bar{K}^{Stat} = \begin{bmatrix} \bar{\mathcal{K}}_1 & & & & 0_{2\times 2(N-2)} \\ & \ddots & & & \\ 0_{2\times 2(i-2)} & & \bar{\mathcal{K}}_i & & 0_{2\times 2(N-i-1)} \\ & & & \ddots & \\ 0_{2\times 2(N-2)} & & & & \bar{\mathcal{K}}_N \end{bmatrix}, \qquad (12a)$$

with $0_{M \times P}$ denoting a zero-filled matrix of $M$ rows by $P$ columns,

$$\bar{\mathcal{K}}_1 = \begin{bmatrix} \mu_1 \Lambda_1^C & \frac{\mu_1 \Lambda_1^C}{2} & -\mu_1 \Lambda_1^C & \frac{\mu_1 \Lambda_1^C}{2} \\ \frac{\mu_1 \Lambda_1^C}{2} & \mu_1\left(\frac{\Lambda_1^C}{4} + \Gamma_1^C + \Gamma_1^B\right) & -\frac{\mu_1 \Lambda_1^C}{2} & \mu_1\left(\frac{\Lambda_1^C}{4} - \Gamma_1^C\right) \end{bmatrix}, \qquad (12b)$$

$$\bar{\mathcal{K}}_i = \begin{bmatrix} -\mu_{i-1}\Lambda_{i-1}^C & -\frac{\mu_{i-1}\Lambda_{i-1}^C}{2} & \mu_{i-1}\Lambda_{i-1}^C + \mu_i \Lambda_i^C & \cdots \\ \frac{\mu_{i-1}\Lambda_{i-1}^C}{2} & \mu_{i-1}\left(\frac{\Lambda_{i-1}^C}{4} - \Gamma_{i-1}^C\right) & \frac{-\mu_{i-1}\Lambda_{i-1}^C + \mu_i \Lambda_i^C}{2} & \cdots \\ & & & \\ & \frac{-\mu_{i-1}\Lambda_{i-1}^C + \mu_i \Lambda_i^C}{2} & -\mu_i \Lambda_i^C & \frac{\mu_i \Lambda_i^C}{2} \\ \cdots & \mu_{i-1}\left(\frac{\Lambda_{i-1}^C}{4} + \Gamma_{i-1}^C\right) + \mu_i\left(\frac{\Lambda_i^C}{4} + \Gamma_i^C + \Gamma_i^B\right) & -\frac{\mu_i \Lambda_i^C}{2} & \mu_i\left(\frac{\Lambda_i^C}{4} - \Gamma_i^C\right) \end{bmatrix}, \qquad (12c)$$

for $2 \leq i \leq N-1$, and,

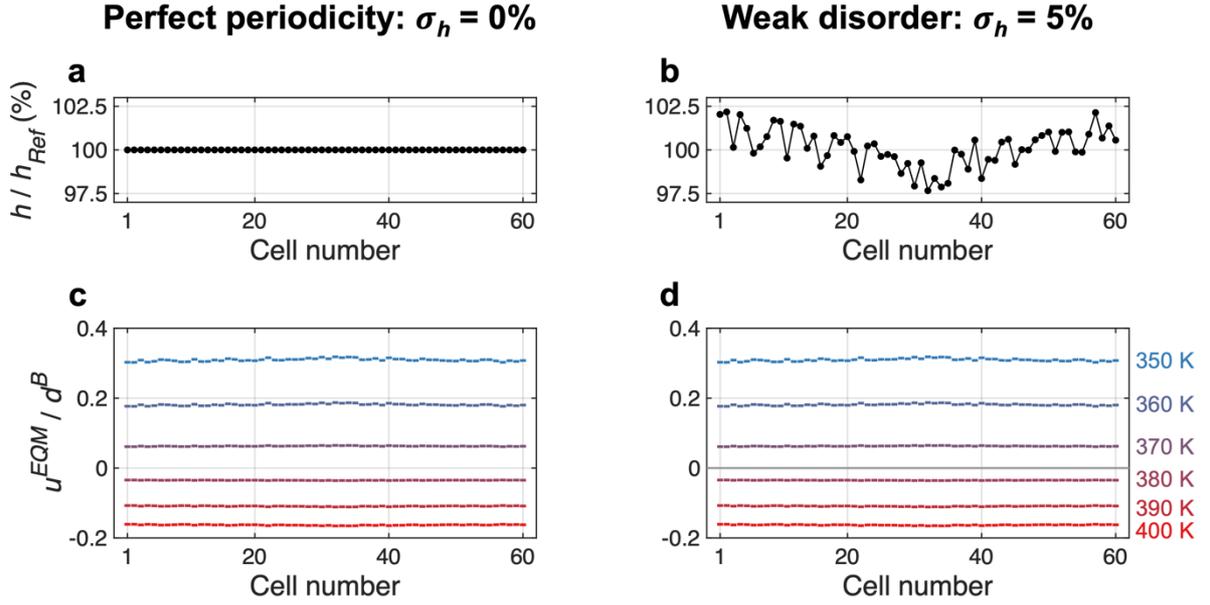

Fig. 3. Effect of weak thickness disorder on the static equilibrium of finite waveguides: (a, b) Thickness profile relative to the reference thickness $h_{Ref}$, and (c, d) static deflections at 400, 390, 380, 370, 360, and 350 K of (a), and (b, d) the weakly ($\sigma_h = 5\%$) disordered 60-cell waveguide ROM of (b); refer to (9) for the definition of the disorder parameter $\sigma_h$; in addition, in (c, d), the thick segments represent the rigid masses of the cells in the ROM of Fig. 1d-e translating and rotating according to the computed equilibria from (11).

$$\bar{\mathcal{K}}_N = \begin{bmatrix} -\mu_{N-1}\Lambda^C_{N-1} & -\frac{\mu_{N-1}\Lambda^C_{N-1}}{2} & \cdots \\ \frac{\mu_{N-1}\Lambda^C_{N-1}}{2} & \mu_{N-1}\left(\frac{\Lambda^C_{N-1}}{4} - \Gamma^C_{N-1}\right) & \\ & & \mu_{N-1}\Lambda^C_{N-1} & -\frac{\mu_{N-1}\Lambda^C_{N-1}}{2} \\ \cdots & -\frac{\mu_{N-1}\Lambda^C_{N-1}}{2} & \mu_{N-1}\left(\frac{\Lambda^C_{N-1}}{4} + \Gamma^C_{N-1}\right) + \mu_N\Gamma^B_N \end{bmatrix}, \quad (12d)$$

where $\Lambda^C_i$, $\Gamma^B_i$, and $\Gamma^C_i$ are defined in (7).

We solve (11) via "*fsolve*" (gradient descent method) in MATLAB®. In the numerical solver, the starting guesses for $\bar{u}^{Eqm}_i$ in (11) correspond to the equilibria of individual cells in (1), see Fig. 2a. We assign the differences $\left(\bar{u}^{Eqm}_{i+1} - \bar{u}^{Eqm}_i\right)$ as starting guesses for $\bar{L}\theta^{Eqm}_i$ for $i \in \{1, 2, \ldots, N-1\}$, and $\bar{L}\theta^{Eqm}_{N-1}$ as the guess for $\bar{L}\theta^{Eqm}_N$. Fig. 3c-d display the computed equilibria of (11) at different temperatures in the *perfectly periodic* and *weakly disordered* waveguides of Fig. 3a-b, respectively. In Fig. 3c-d, each segment (thick dashed line) represents a cell in the ROM of Fig. 1d-e translated and rotated according to the equilibrium of (11).

In Fig. 3c, the cells at equilibrium undergo the same translational deflections without rotation, which results from the perfect periodicity of the waveguide of Fig. 3a. For instance, the weakly-disordered waveguide of Fig. 3b attains equilibrium with different cell translations and rotations, as shown in Fig. 3d. For both waveguides of Figs. 3c-d, the cooling increases the cells' baseline translational deflections going from negative to positive values between 400 K and 350 K(like the individual cell in Fig. 2a). Moreover, each 10 K of cooling induces larger deflections at lower temperatures in Figs. 3c-d, which mirrors the buckling susceptibility observed in Fig. 2a.

At this point, we linearize the dynamics around the calculated equilibria to study the transmission in the finite waveguides for varying temperatures. Using the perturbation coordinates in (5), the linearized equations yield,

$$\overline{M}\frac{d^2\overline{q}}{dt^2} + \underbrace{(\overline{K}^{Buck} + \overline{K}^{Stat})}_{\overline{K}} \overline{q} = \mathbf{0}, \tag{13}$$

where $\overline{q} = \{\overline{v}_1, \overline{h}_1, \ldots, \overline{v}_i, \overline{h}_i, \ldots, \overline{v}_N, \overline{h}_N\}^T$, $\overline{K}^{Stat}$ is defined in (12), $\overline{K}^{Buck}$ corresponds to the $2N \times 2N$ matrix whose diagonal contains the linear stiffnesses of the buckling forces,

$$\overline{K}^{Buck} = \text{diag}(\mu_1 \Lambda_1^{Buck}, 0, \ldots, \mu_i \Lambda_i^{Buck}, 0, \ldots, \mu_N \Lambda_N^{Buck}, 0), \tag{14}$$

and $\overline{M}$ denotes the nondimensional mass matrix expressed as:

$$\overline{M} = \begin{bmatrix} \overline{\mathcal{M}}_1 & & & & 0_{2\times 2(N-1)} \\ & \ddots & & & \\ 0_{2\times 2(i-1)} & & \overline{\mathcal{M}}_i & & 0_{2\times 2(N-i)} \\ & & & \ddots & \\ 0_{2\times 2(N-1)} & & & & \overline{\mathcal{M}}_N \end{bmatrix}, \tag{15}$$

where $\overline{\mathcal{M}}_i = \mu_i \begin{bmatrix} 1 & 0 \\ 0 & \chi \end{bmatrix}$ for $i \in \{1,2,3,\ldots,N\}$. In this work, we solve for the acoustics of the waveguides by direct integration of (13) using MATLAB® "*ode45*" function.

To this end, we consider the solution of (13) subject to a nonzero initial translational velocity at cell 1 and all other initial conditions set to zero:

$$\mathbf{q_0} \stackrel{\text{def}}{=} \overline{q}(t=0) = \mathbf{0} \text{ and } \dot{\mathbf{q}}_0 \stackrel{\text{def}}{=} \frac{d\overline{q}}{dt}(t=0) = \begin{Bmatrix} 10^{-3}\mu_1 \\ 0 \\ \mathbf{0}_{2(N-1)\times 1} \end{Bmatrix}. \tag{16}$$

The initial conditions (16) induce a motion that propagates in the waveguides as depicted in Fig. 4-5 and the supplemental Video.S1. This motion enables the study of wave transmission in the considered finite waveguides. In Figs. 4-5 and supplemental Video.S1, we consider the responses of the waveguides of Figs. 3a-b at 390 K, to address the effect of weak disorder at a fixed temperature corresponding to (relatively) broad passbands I and II based on the Bloch modes of Figs. 2b-c.

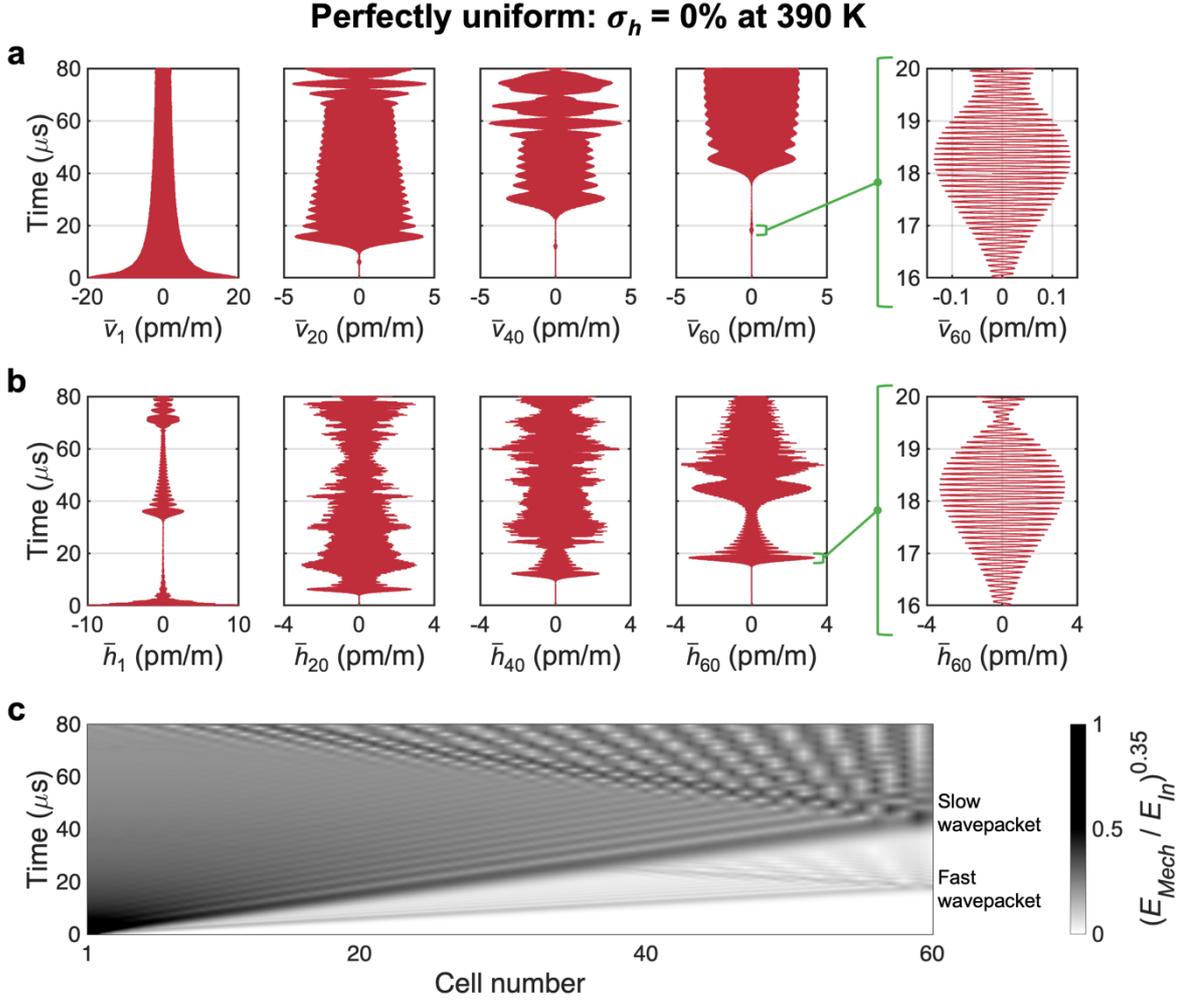

Fig. 4. Elastic wave transmission through the *perfectly periodic finite* waveguide of Fig. 3a at 390 K: Temporal responses in terms of (a) the translations $\bar{v}_n$ and (b) the rotational $\bar{h}_n$ perturbation coordinates of (from left to right) cell $n \in \{1, 20, 40, 60\}$ due to the initial conditions in (16) (the rightmost plots in (a, b) show zoomed-in views of the responses of cell 60 in the period [16, 20] $\mu$s); (c) spatiotemporal evolution of the normalized mechanical energy $E_i^{Mech}/E^{In}$ of (17) for $i \in \{1, 2, …, 60\}$ in the ROM waveguide – the wavepacket labels highlight the instants at which the fast and slow wavepackets reach cell 60 and reflect.

Figs. 4a-b show the translational and rotational time-responses, respectively, for wavepackets propagating through the waveguide of Fig. 3a at a temperature of 390 K. We observe that the initial velocity imposed at cell 1 by (16) provokes a wavepacket that propagates to the last cell (with index 60) (cf. also supplemental Video.S1). The propagating front of this wave consists mainly of two distinct wavepackets with different wave speeds, namely, a "fast" wavepacket reaching cell 60 within < 20 $\mu$s, and a "slow" wavepacket reaching that boundary cell within ~40 $\mu$s. Fig. 4a shows the fast wavepacket characterized by low translational

amplitudes compared to the slow wavepacket. Fig. 4b shows both wavepackets possessing similar rotational amplitudes. The slow and fast wavepackets in the *finite* waveguide are analogous to waves transmitted in passbands I and II, respectively, of the *infinite perfectly periodic* waveguide (cf. Figs. 2b-c). We conclude that at the temperature considered, the finite waveguide supports the propagation of *spatially extended* wavepackets with frequency-wavenumber contents lying inside passbands predicted in the waveguide of infinite extent. To visualize the waves propagation over every cell, in Fig. 4c, we plot the spatiotemporal normalized energy evolution in the corresponding *perfectly periodic* 60-cell waveguide. In particular, we depict the contour plot of the normalized mechanical energy $E_i^{Mech}/E^{In}$ of each cell $i \in \{1, 2, ..., N\}$ defined by,

$$\frac{E_i^{Mech}}{E^{In}} = \left\{ \mu_i \left(\frac{d\bar{v}_i}{dt}\right)^2 + \mu_i \chi \left(\frac{d\bar{h}_i}{dt}\right)^2 + \Lambda_i^{Buck} \bar{v}_i^2 + \Gamma_i^B \bar{v}_i^2 \right.$$
$$+ \frac{1}{2} \left[ \Lambda_{i-1}^C (\bar{v}_i - \bar{v}_{i-1})^2 + \Lambda_i^C (\bar{v}_{i+1} - \bar{v}_i)^2 + \Gamma_{i-1}^C (\bar{h}_i - \bar{h}_{i-1})^2 \right. \quad (17)$$
$$\left. \left. + \Gamma_i^C (\bar{h}_{i+1} - \bar{h}_i)^2 \right] \right\} / (\dot{\boldsymbol{q}}_\mathbf{0}^T \overline{M} \dot{\boldsymbol{q}}_\mathbf{0} + \boldsymbol{q}_\mathbf{0}^T \overline{K} \boldsymbol{q}_\mathbf{0}),$$

with $\Lambda_0^C = \Lambda_{N+1}^C = \Gamma_0^C = \Gamma_{N+1}^C = 0$. Fig. 4c demonstrates the two waves propagating in the primary front, reaching the waveguide boundary at cell 60. Moreover, Fig. 4c shows the slow wavepacket transmitting a significantly larger portion of the input energy since it mainly corresponds to the translational motion (cf. Fig. 4a) that is more effectively excited via the initital conditions in (16).

For comparison, in Fig. 5 we depict the corresponding wave transmission in the ($\sigma_h = 5\%$) ***weakly disordered*** waveguide at 390 K, forced by the same excitation in (16). A drastically different acoustics are observed for the disordered system. We observe that only the early (fast) wavepacket propagates to cell 60 in the presence of disorder. Moreover, as Fig. 5c shows, the slow wavepacket (which carries the major portion of the available energy) becomes spatially localized in the first 30 cells, a result which indicates that the *weakly disordered* waveguide at 390 K cannot transmit the slow wavepacket corresponding to passband I of Figs. 2b-c. This transmission loss for wavepackets in passband I is in full agreement with the experimental findings reported in [29] for a similar waveguide (and summarized in Figs. 1a-c). Therefore, the ROM developed in this work captures the experimental transmission loss mediated by buckling and provides conclusive proof regarding the important role that structural disorder plays for the transmission loss in buckled waveguides at certain temperature ranges.

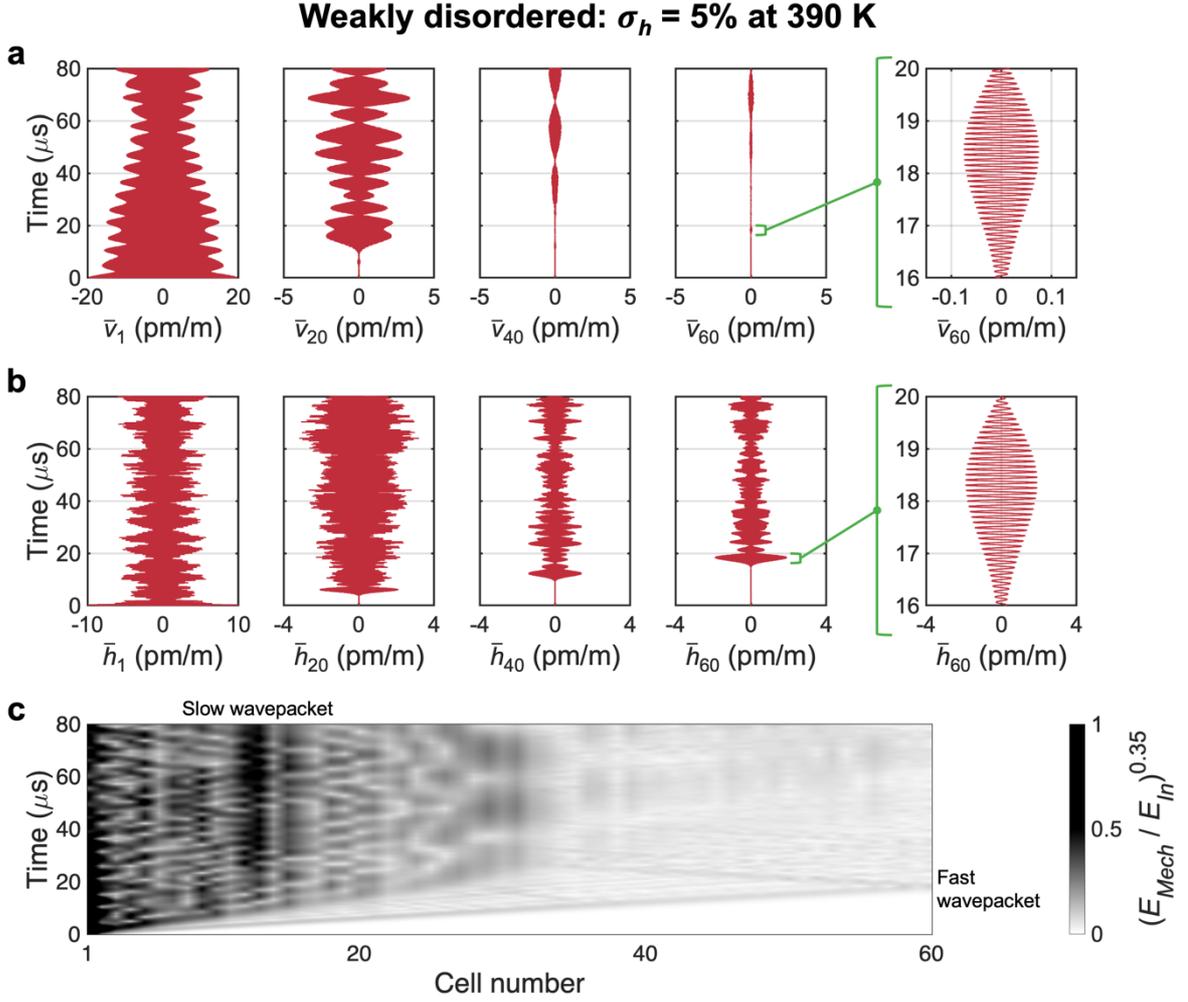

Fig. 5. Elastic wave transmission through the *weakly disordered finite* waveguide of Fig. 3c at 390 K: Temporal responses in terms of (a) the translations $\bar{v}_n$ and (b) the rotational $\bar{h}_n$ perturbation coordinates of (from left to right) cell $n \in \{1, 20, 40, 60\}$ due to the initial conditions in (16) (the rightmost plots in (a, b) show zoomed-in views of the responses of cell 60 in the period [16, 20] $\mu$s; (c) spatiotemporal evolution of the normalized mechanical energy $E_i^{Mech}/E^{In}$ of (17) for $i \in \{1, 2, \ldots, 60\}$ in the ROM waveguide – the "Fast wavepacket" label highlights the instant at which the fast wavepacket reaches cell 60 and reflect; the "Slow wavepacket" label highlights the cells where the slow wavepacket is confined.

This transmission loss mechanism is also observed by the finite element model (FEM) of the waveguide studied in [29], further illustrated in supplemental Video.S2 presenting the time-series deformations of the centerlines of two waveguides based on the COMSOL simulations of [29]. In particular, we consider two identical 20-cell *weakly disordered* waveguides at -20 K (i.e., far from critical buckling) and -120 K (close to critical buckling), respectively. In supplemental Video.S2, we assess the waveguides capacity to transmit propagating wavepackets with frequency contents inside passband I of their corresponding infinite waveguide (i.e. the passbands based on the Bloch modes of the constitutive unit cell). These FEM simulations show that an elastic wavepacket can propagate only in *the weakly disordered* waveguide far from critical buckling (cf. [29] for the FEM geometry and methods). Hence, with the ROM developed in this work, we confirm that buckling-induced transmission loss for waves in passband I is associated with weak disorder and thermo-elastic effects, confirming the experimental and FEM results reported in [29].

### V. Frequency transmission in the finite waveguides

To further study the frequency transmission in the considered waveguides at 390 K, Fig. 6 presents the amplitude of the Fast Fourier transforms (FFT) of the displacements at selected cells subject to the initial conditions (16). For comparison, we overlay the FFT plots on top of the Bloch modes' passbands of the infinite waveguide (cf. Figs. 2a-c) and the modal frequencies $\omega^{Mode} = \sqrt{\Lambda^{Mode}}$ of the finite waveguide calculated by the eigenvalue problem,

$$(-\Lambda^{Mode}\overline{M} + \overline{K})\,\boldsymbol{\Phi} = \boldsymbol{0}, \qquad (18)$$

where $\boldsymbol{\Phi}$ denotes the mode shape vector. For clarity, we do not show in Fig. 6 the modal frequencies that lie within the passbands.

Figs. 6a-b illustrate that all modal frequencies of the *perfectly periodic* finite waveguide are inside the passbands (there exist 60 modes in each passband); whereas Figs. 6c-d shows certain modes of the ($\sigma_h = 5\%$) *weakly disordered* finite waveguide are lying outside these passbands, i.e., in stopbands. Hence, the Bloch modes of the infinite waveguide constitute a *perfect estimator* of wave transmission only in the *perfectly periodic* finite waveguide.

In Figs. 6a-b, the *perfectly periodic* finite waveguide corresponds to strong translational and rotational responses at cell 60, respectively, only when their frequency contents are inside the passbands. Outside of the passbands, however, the frequency responses of the responses at cell 60 are minimal compared to the response of the excited cell 1, indicating a lack of transmission throughout the waveguide (i.e., stopband). In Figs. 6c-d, cell 60 of the ($\sigma_h = 5\%$) *weakly disordered* finite waveguide admits weak responses compared to the response of cell 1 in the Bloch modes defining passband I. These frequency responses verify that the transmission loss shown in Fig. 5 corresponds to the transmission loss of wavepackets inside passband I. Notably, concerning wavepackets initiated in passband II (cf. Fig. 6d), the rotational response of cell 60 compares in magnitude to cell 1 due to the persistence of the transmission in this passband, as previously illustrated in Fig. 5.

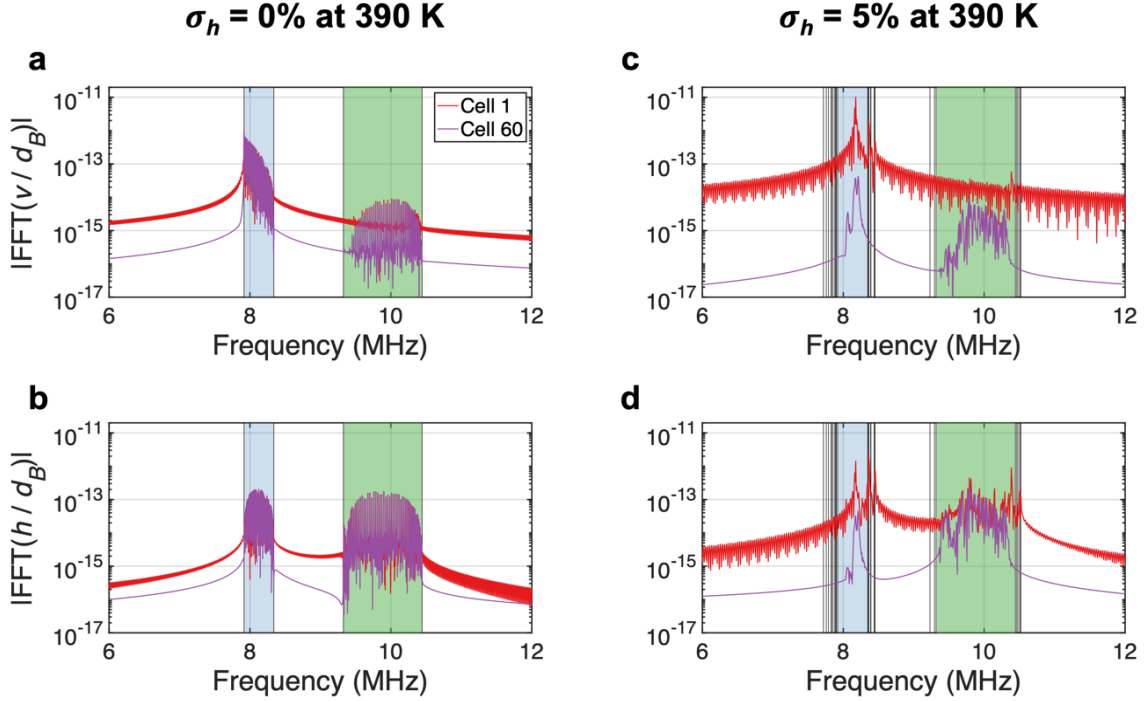

Fig. 6. Effect of weak thickness disorder on the frequency responses of finite waveguides: Fast Fourier Transforms (FFT) of the (a,c) translational and (b,d) rotational temporal responses of of cells 1 (red line), and 60 (purple line), at 390 K of the *perfectly periodic* 60-cell waveguide in (a,b), and the *weakly disordered* 60-cell waveguide in (c, d) (refer to the ROM of Figs. 3a,c); the blue and green shaded background regions correspond to the frequencies of passbands I and II of the corresponding perfectly periodic waveguide at 390K, respectively(cf. Figs. 2b-c), whereas the black vertical lines denote the modal frequencies of the finite waveguides calculated using (18) whose values are not inside the Bloch modes passbands.

Similar performance to the results of Figs. 4-6 is observed for different temperatures between 400 and 350 K, as shown in Fig. 7. In particular, the *perfectly periodic* finite waveguide of Fig. 3a does not lose transmission for the considered temperatures; the *weakly disordered* finite waveguide of Fig. 3c cannot transmit waves in passband I between 390 and 352 K.

In Figs. 7-8, we summarize the results for all measured temperatures by considering the frequency contents of the wavepackets transmitting throughout the spatial extent of the *perfectly periodic* or *weakly disordered* waveguides. We relate the frequency transmission to the nondimensional kinetic energy $\bar{E}_i^{Kin}(t)$ attained by the cell of index $i \in \{1, 2, \ldots, N\}$ and numerically approximated by,

$$\bar{E}_i^{Kin}(t) \cong \sum_{m=1}^{n_{FFT}} \tilde{E}_{i,m} \cos(\tilde{\omega}_m t + \tilde{\theta}_m), \qquad (19a)$$

$$\text{with } \tilde{E}_{i,m} = \frac{1}{2}\mu_i \widetilde{\omega}_m{}^2 \left[\tilde{v}_{i,m}{}^2 + \chi \tilde{h}_{i,m}{}^2\right], \tag{19b}$$

where we denote by $n_{FFT}$ the output number of FFT sampled frequencies $\widetilde{\omega}_m \geq 0$ rad/s with phase shifts $\tilde{\theta}_m$, and $\tilde{v}_{i,m}$ and $\tilde{h}_{i,m}$ the FFT amplitudes of $\bar{v}_i(t)$ and $\bar{h}_i(t)$, respectively, at frequencies $\widetilde{\omega}_m$ for $m \in \{1, 2, \ldots, n_{FFT}\}$. Equation (19) assumes that the FFT of both $\bar{v}_i(t)$ and $\bar{h}_i(t)$ have identical phase shifts $\tilde{\theta}_m$ at $\widetilde{\omega}_m$ leading to $\bar{q}_i(t) \cong \sum_{m=1}^{n_{FFT}} \tilde{q}_{i,m} \cos(\widetilde{\omega}_m t + \tilde{\theta}_m)$ for $q \in \{v, h\}$.

In Figs. 7-8, the contour plots depict the $\tilde{E}_{i,m}$ normalized by max $\tilde{E}_{j,p}$ over all $j \in \{2, 3, \ldots, N\}$ and all $p \in \{1, 2, \ldots, n_{FFT}\}$ of the respective response. For better visualization, we coerce the contour values to 0 and 1 if they fall below the minimum threshold $\mathcal{E}_{Ths} \leq 2 \times 10^{-3}$, and above the saturation limit $\mathcal{E}_{Sat} \geq 0.5$, respectively. In essence, in Figs. 7-8, we assess the binary behavior of the considered waveguide by checking whether the energy can transmit to cell $i$ at frequency $\widetilde{\omega}_m$ under the studied temperature and disorder conditions.

In particular, we investigate the acoustics of three waveguides with disorders $\sigma_h \in \{0\%, 2.5\%, 5\%\}$ as defined in (9); the corresponding thickness profiles of the waveguides are provided in Fig. 3a, the supplemental Video.S3, and Fig. 3c, respectively. Moreover, to efficiently excite passband II in addition to passband I, we modify in this section the initial conditions in (16) into:

$$\boldsymbol{q_0} \stackrel{\text{def}}{=} \bar{\boldsymbol{q}}(t = 0 \text{ s}) = \boldsymbol{0} \text{ and } \dot{\boldsymbol{q}}_0 \stackrel{\text{def}}{=} \frac{d\bar{\boldsymbol{q}}}{dt}(t = 0 \text{ s}) = \begin{Bmatrix} 10^{-3}\mu_1 \\ 10^{-3}\mu_1\sqrt{\chi} \\ \boldsymbol{0}_{2(N-1)\times 1} \end{Bmatrix}. \tag{20}$$

Note that by the new initial conditions (20), we deliver the same initial kinetic energy for both translational and rotational coordinates of cell 1. We use in (20) the same value of initial kinetic energy delivered in (16) in order to provide a fair analogy to the previous results.

The plots of Figs. 7a-e display the wave transmission in the three finite waveguides at 400, 390, 370, 353, and 350 K, respectively. For all these temperatures, the *perfectly periodic* finite waveguide ($\sigma_h = 0\%$) transmits energy to the last cell 60, with frequency contents in both passbands I and II – cf. Figs. 7(i). This transmission for all temperatures is unique for the *perfectly periodic* waveguide and does not occur in the *weakly disordered* finite waveguides considered in Figs. 7(ii)-(iii). For example, we observe transmission loss of waves with frequency content in passband I for the waveguide with $\sigma_h = 2.5\%$ at 370 K in Fig. 7c(ii), and for the waveguide with $\sigma_h = 5\%$ at 390, 370, and 353 K in Figs. 7b-d(iii), respectively. Hence, larger disorders correspond to a more extended range of temperatures with transmission loss in passband I.

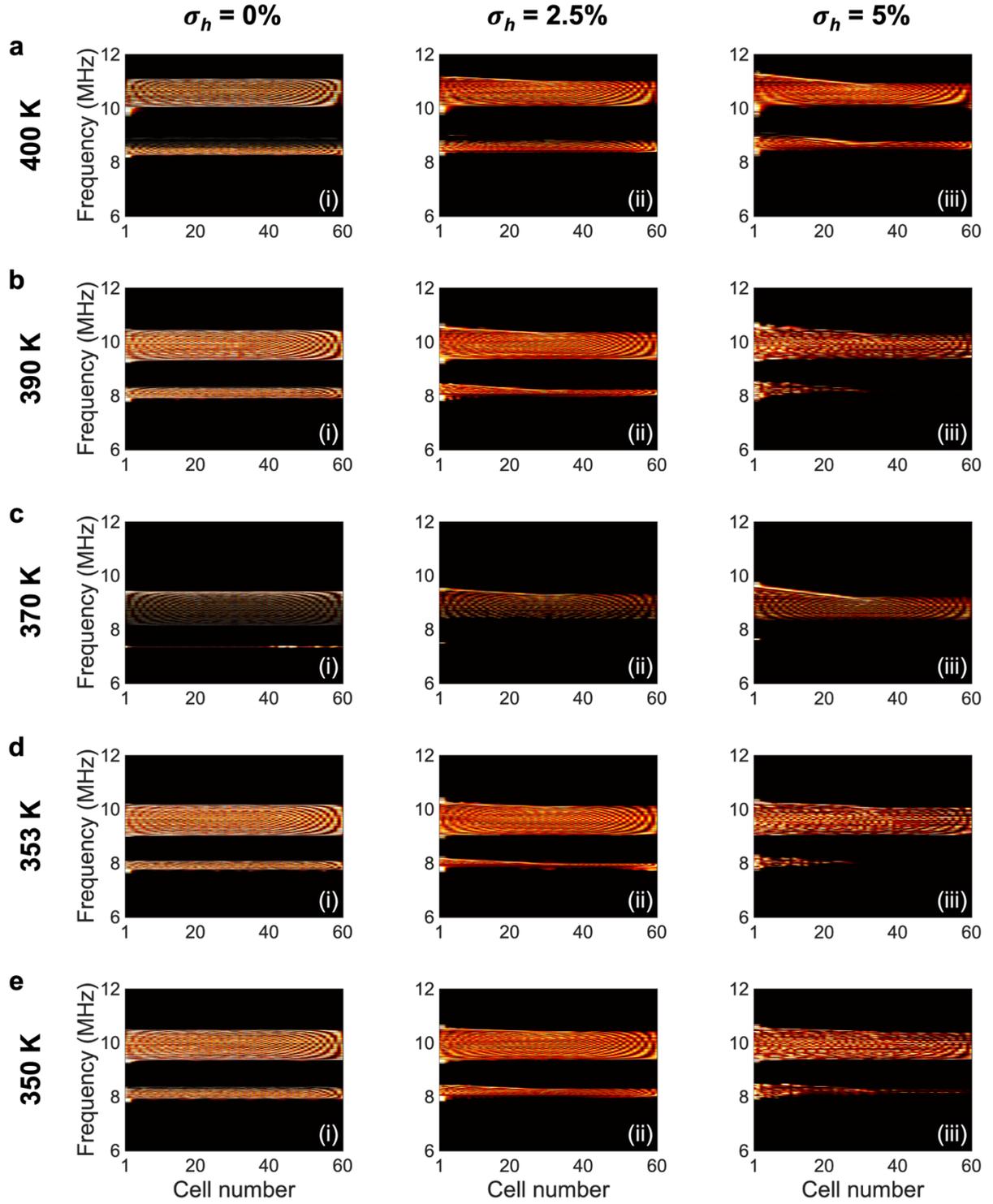

Fig. 7. Effects of thickness disorder and thermal buckling on the frequency content of transmitted waves through finite waveguides: Contour plots of normalized transmitted energy vs. cell number and wave frequency at (a) 400 K, (b) 390 K, (c) 370 K, (d) 353 K, and (e) 350 K through the 60-cells waveguides with thickness disorders $\sigma_h$ of (i) 0%, (ii) 2.5%, and (iii) 5%. ; black and yellow colors correspond to the limiting values of 0 and 1, respectively.

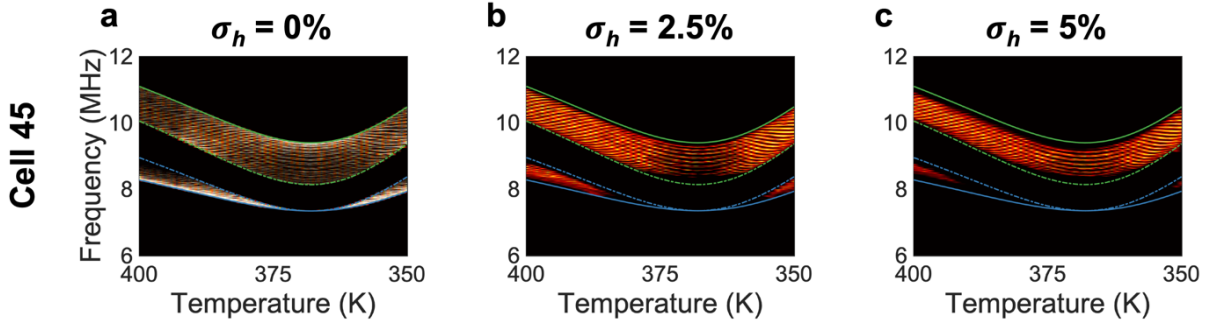

Fig. 8. Effects of thickness disorder and thermal buckling on the frequency content of transmitted waves reaching 75% into the finite waveguides: Contour plots of the transmitted energy to cell 45 vs. the temperature and the wave frequency in the 60-cell waveguides with thickness disorders $\sigma_h$ of (a) 0%, (b) 2.5%, and (c) 5%; the displayed values are extracted from cell 45 similarly to the plots of Fig. 7 but over the temperature domain (350K, 400K); the black and yellow colors correspond to the limiting values of 0 and 1 for the normalized energy, respectively, the blue lines to the frequency extrema of passbands I and II, respectively, cf. Fig. 2c, and the solid/dashed lines to the filled and open circles in Fig. 2c, respectively.

Moreover, we notice that the frequency width of passband I is smaller in the transmitting scenarios of the *weakly disordered* waveguides compared to the *perfectly periodic* waveguide. This width-narrowing of passband I accompanies a similar narrowing of passband II in the *weakly disordered* waveguides of Figs. 7(ii)-(iii). However, contrary to passband I, the *weakly disordered* waveguides keep transmitting energy to the last cell 60 in passband II at all temperatures considered, as depicted in Figs. 7. Therefore, we conclude that passband II is less susceptible to structural disorder than passband I, which fully agrees with what was experimentally witnessed in [29]. Hence, the ROM developed herein accurately captures these acoustic aspects of the phononic lattice under investigation.

To further clarify the dependency of wave transmission on temperature, in Figs. 8a-c, we study the frequency content of transmitted waves reaching cell 45 in the three finite waveguides with disorder $\sigma_h \in \{0\%, 2.5\%, 5\%\}$, respectively. On top of the finite waveguides results, we overlay the corresponding passbands of the infinite waveguides, cf. Fig. 2c. The results in Fig. 8a prove that, at all the considered temperatures, the wave transmission in the *perfectly periodic* finite waveguide have frequency contents solely inside the passbands. Thus, the passbands of the infinite waveguide perfectly estimate the wave transmission in the *perfectly periodic* finite waveguide at all temperatures, as concluded in the previous section from Figs. 6a-b at 390 K.

However, this perfect estimation does not hold for the wave transmission in the *weakly disordered* finite waveguides considered in Figs. 8b-c, especially around the critical-buckling temperature (i.e., 370 K). Although, as discussed previously, there is a band-narrowing of both passbands I and II, wave transmission loss is only realized for passband I, cf. Figs. 8b-c. This behavior confirms that passband I is more susceptible to disorders than passband II, confirming the analogous conclusion illustrated in Fig. 1c of the experimental work in [29]. Lastly, by comparing Figs. 8b to 8c, we deduce that stronger disorders result in more severe wave transmission losses with thermal-mediated buckling.

## VI. Buckling-induced localized modes

To explain the causes of transmission loss due to disorder, we plot in Figs. 9-10 the spatial distributions (modeshapes) of two modes of the finite waveguides, specifically modes 30 and 90, at 400, 390, 370, 353, and 350 K. Thick dash lines represent the normalized translational and rotational deformations of individual cells calculated by the eigenvector $\Phi$ of (18) at the considered temperature and waveguide. For better visualization, we join the cells with cubic splines depicted as thin lines to imitate the continuous deformation of the waveguide's centerline. Mode 30 considered in Fig. 9 is inside passband I of the infinite *perfectly periodic* waveguide (which contains also the first 60 modes of the *perfectly periodic* finite waveguide with no disorder), whereas mode 90 in Fig. 10 is located inside passband II (which also contains modes 61 to 120 of the corresponding *perfectly periodic* finite waveguide with no disorder).

Figs. 9a-e(i) show that mode 30 of the *perfectly periodic* finite waveguide deforms with comparable amplitudes over the entire spatial extent of the system, i.e., from cell 1 (the excited cell) up to cell 60, at all the considered temperatures. Such modes with spatially extended amplitude distributions over the entire waveguide are called *"extended modes"* that are necessary for wave propagation in the finite waveguides. For instance, this type of *extended modes* enable energy initially applied to cell 1 to appreciably deform the remaining cells resulting in detectable mechanical energy propagation through the entire spatial length of the waveguide. The *extended modes* in Figs. 9a-e(i) verifies the persistence of wave transmission via passband I of the *perfectly periodic* finite waveguide as depicted in Figs. 7a-e(i) and 8a *for all temperatures*, even very close to the critical-buckling temperature.

Conversely, in the *weakly disordered* finite waveguides not all modes are extended for all temperatures. For example, for disorder level $\sigma_h = 5\%$, mode 30 is an extended mode only at 400 and 350 K, respectively – cf. Figs. 9(ii)a,e. This extended shape directly affects wave transmission through the waveguide: A nonzero initial deformation of cell 1 due to the applied excitation deforms the cells throughout the waveguide as shown by the modeshapes of Figs. 9(ii)a and 9(ii)e, allowing energy to propagate throughout the waveguide (cf. Figs. 7(iii)a,e and 8c). However, cell 1 exhibits minimal vibrations in Figs. 9(ii)b-d at 390, 370, and 353 K, respectively, justifying the inability to transmit energy by mode 30 in Figs. 7(iii)b-d and 8c. The modeshapes in Figs. 9(ii)b-d are "localized modes" where large deformations are confined only locally without extending over the entire waveguide (like in extended modes).

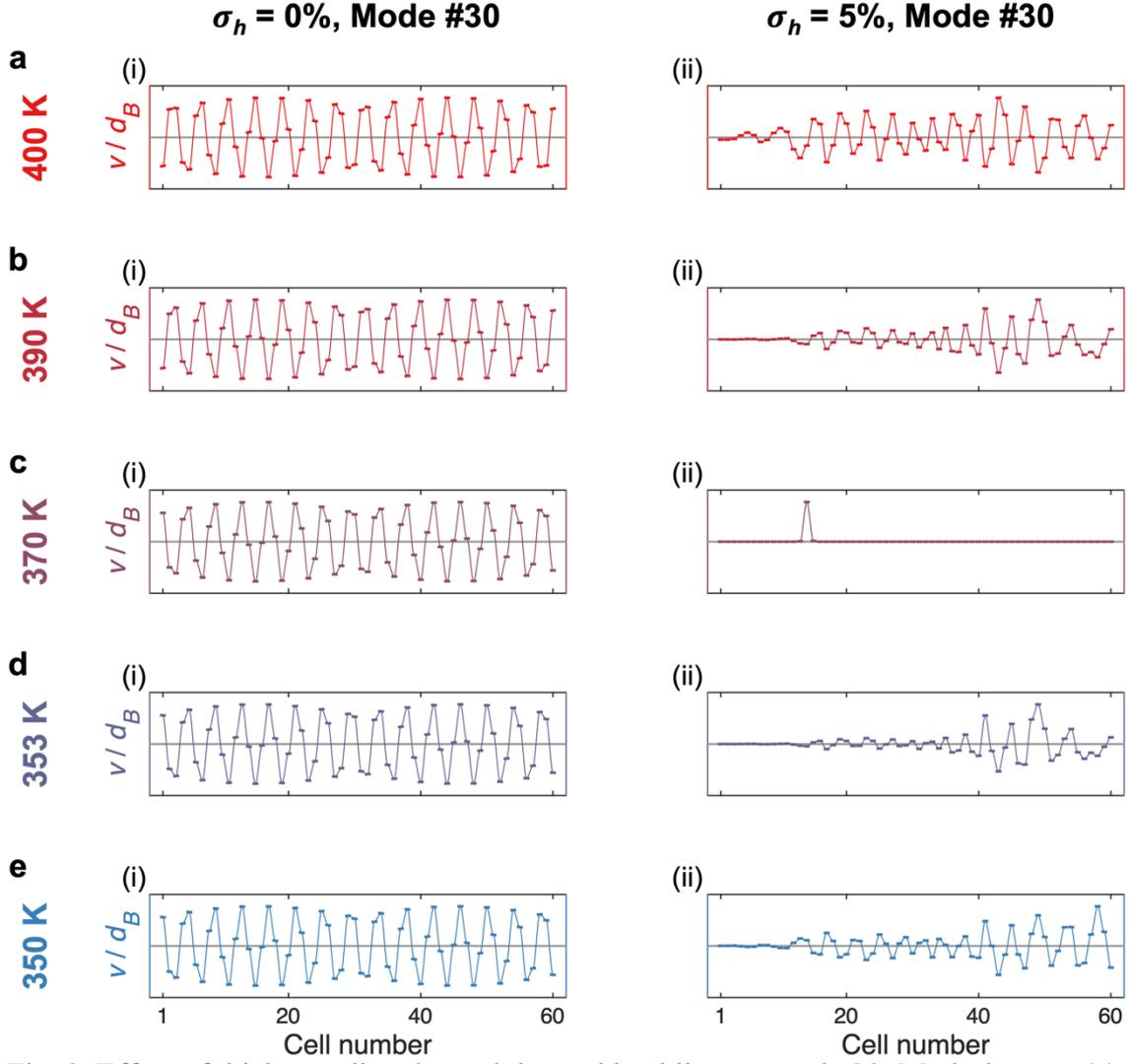

Fig. 9. Effect of thickness disorder and thermal buckling on mode 30: Modeshape at (a) 400 K, (b) 390 K, (c) 370 K, (d) 353 K, and (e) 350 K in (i) the *perfectly periodic* finite waveguide (where mode 30 is in passband I), and (ii) the ($\sigma_h = 5\%$) *weakly disordered* finite waveguide.

*Localized modes* characterize aperiodic/disordered structures because *extended modes* necessitate the periodicity between the constitutive cells in the waveguides. In other words, cells of similar geometry and material configurations form a *periodic* structure with cells of similar (isolated) modal frequencies, which we refer to as *modal periodicity*. This *modal periodicity* is necessary to form the *extended modes* that enable transmission throughout the structure. In this work, we conjecture that the *modal periodicity* breaks in the *weakly disordered* waveguides under the effect of buckling because the cells in the waveguides develop different (isolated) modal frequencies in passband I due buckling-induced changes in the grounding and coupling stiffnesses (cf. [41]). The buckling-induced differences in modal frequencies lead to *localized modes*, like in Fig. 9(ii)b-d, inhibiting energy transmission throughout the waveguides.

As a general conclusion, buckling and thermoelastic effects lead to shifts of modeshapes in the frequency domain due to disorder, which, in turn, "transforms" certain modes from *extended* to *localized*. Therefore, *thermal effects and buckling magnify the "modal" disorder in the disordered waveguides, leading to energy localization and confinement, similar to Anderson localization* [42]. Due to this effect, many modes between #1 and #60 (inside passband I of the *perfectly periodic* finite waveguide) become localized in the *weakly disordered* waveguides, as seen at 390 K in the supplemental Video.S3. Indeed, at 390 K, all the leading 60 modes are localized in the waveguide with $\sigma_h = 5\%$, prohibiting passband I transmission, cf. Figs. 7(iii)b and 8c. In addition, supplemental Video.S3 demonstrates the existence of some extended modes between mode #1 and #60 in the waveguide with $\sigma_h = 2.5\%$ at 390 K, which explains the observed passband I transmission at 390 K of Figs. 7(ii)b and 8b. Lastly, in Supplemental Video.S3, all the leading 60 modes of the *perfectly periodic* waveguide are extended modes, resulting in broader passband I at 390 K than the waveguide with $\sigma_h = 2.5\%$ – compare Figs. 7b(i) to 7b(ii). Note that all 60 leading modes of the *perfectly periodic* waveguide are extended modes even at the critical-buckling temperature of 370 K, as shown in supplemental Video.S4.

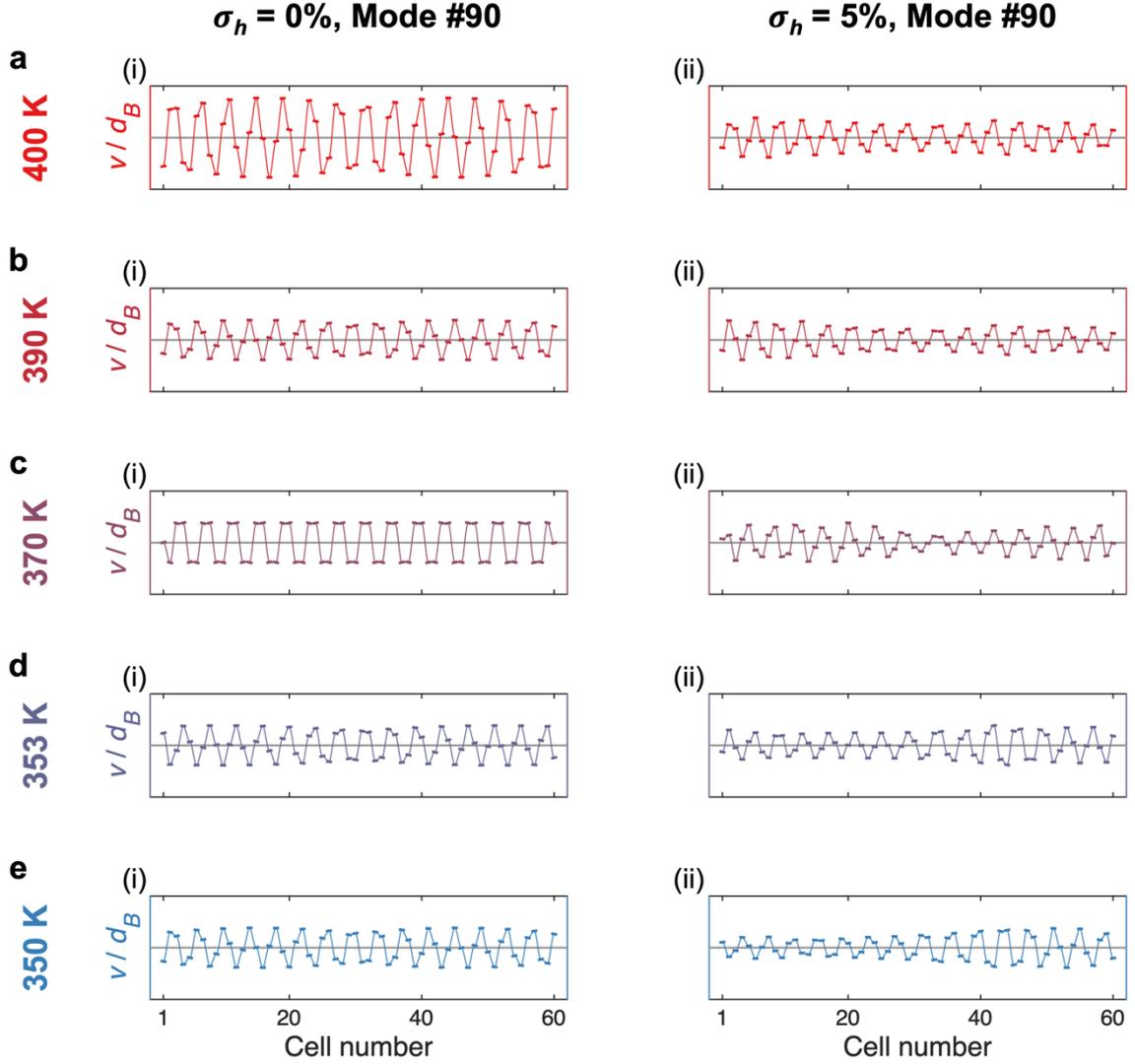

Fig. 10. Effect of thickness disorder and thermal buckling on the mode 90: Modeshape at (a) 400 K, (b) 390 K, (c) 370 K, (d) 353 K, and (e) 350 K; in (i) the *perfectly periodic* finite waveguide (where mode 30 is in passband II), and (ii) the ($\sigma_h = 5\%$) *weakly disordered* finite waveguide.

The buckling-induced localization does not occur in mode 90 of Fig. 10, not even in the *weakly disordered* waveguide of $\sigma_h$ = 5%. Therefore, mode 90 remains an extended mode at all temperatures and weak disorders, enabling the propagation of a wave at the modal frequency of mode 90. Most modes between 61 and 120 possess similar extended modeshapes, as illustrated in supplemental Video.S5 and Video.S6 at 390 K and the critical-buckling temperature of 370 K, respectively. In addition, as expected, all modes between #61 and #120 are extended in the *perfectly periodic* finite waveguide at all temperatures. In contrast, some modes away from the median mode 90 are localized in the *weakly disordered* finite waveguides, resulting in the narrowing of passband II in Figs. 8b-c. Supplemental Video.S5 and Video.S6 show fewer localized modes for weaker disorders (i.e., $\sigma_h$ of 2.5% vs. 5%), verifying the wider passband II of Fig. 8b compared to Fig. 8c.

## VII. Conclusions

We developed a reduced-order model (ROM) for on-chip phononic waveguides made from coupled drumhead resonators. In particular, we investigated the effect of thermal-induced buckling on eliminating transmission over a low-frequency passband in *weakly disordered* waveguides. The considered disorders are very small and typically result from fabrication errors. We show that buckling magnifies the effect of *weak geometric aperiodicity* by amplifying *the modal disorders* between constitutive cells of the waveguide. The resulting effective aperiodicity yields transmission loss in the first passband of the waveguide due to spatial localization of subsets of modes, similar to Anderson localization. This localization is ineffective for the higher-frequency passbands of the disordered waveguides, which support robust transmission to disorder and buckling.

Notably, the developed ROM can capture the dynamics of a two-passband waveguide under thermal buckling. We adopt the buckling model from the experimental results in [41] by introducing the thermal expansion of the plate microstructure of the individual cells (undergoing stretching and bending) and the fabrication-residual stresses. Moreover, we control the level of buckling in the ROM waveguide by assigning different temperatures. We study the transmission as a function of temperature by considering the Bloch modes and the free response of the *perfectly periodic* finite waveguides. Hence, we present a method to relate the free response to the frequency content of transmitted waves in the waveguide, which saves computational effort compared to simulating the frequency response of the ROM.

The present study highlights the important role of validated ROMs in the design of phononic or acoustic waveguides that undergo buckling phase transitions. In these cases, Bloch mode analysis fails to capture the experimental results even for *weakly disordered* finite waveguides. We highlight the fact that transmission in finite waveguides is achieved via *extended modes* of the waveguides, whereas the existence of *localized modes* inhibits wave transmission, as energy becomes spatially confined and does not transmit throughout the extent of the waveguide. These results are fully captured by the developed ROM, which offers a reliable and robust alternative predictive design tool for on-chip phononic waveguides, compared to experimental and/or finite element computational methods which are not as versatile or computationally inexpensive.

# VIII. Acknowledgment

This work was supported in part by NSF Emerging Frontiers in Research and Innovation (EFRI) Grant 1741565. This support is greatly acknowledged by the authors.

# IX. References

[1] R. H. Olsson and I. El-Kady, "Microfabricated phononic crystal devices and applications," *Measurement science and technology,* vol. 20, no. 1, p. 012002, 2008.

[2] R. Aigner, "Volume manufacturing of BAW-filters in a CMOS fab," *Second International Symposium on Acoustic Wave Devices for Future Mobile Communication Systems,* pp. 3-5, 2004.

[3] K. M. Lakin, "Thin film resonator technology," *EEE Transactions on Ultrasonics, Ferroelectrics, and Frequency Control,* vol. 52, no. 5, pp. 707-716, 2005.

[4] S. AIP Mohammadi and A. Adibi, "On chip complex signal processing devices using coupled phononic crystal slab resonators and waveguides," *AIP Advances,* vol. 1, no. 4, p. 041903, 2011.

[5] M. B. Zanjani, A. R. Davoyan, A. M. Mahmoud, N. Engheta and J. R. Lukes, "One-way phonon isolation in acoustic waveguides," *Applied Physics Letters,* vol. 104, no. 8, p. 081905, 2014.

[6] H. Zhu and F. Semperlotti, "Double-zero-index structural phononic waveguides," *Physical Review Applied,* vol. 8, no. 6, p. 064031, 2017.

[7] A. Buhrdorf, O. Ahrens and J. Binder, "Capacitive micromachined ultrasonic transducers and their application," *2001 IEEE Ultrasonics Symposium. Proceedings. An International Symposium (Cat. No. 01CH37263),* vol. 2, pp. 933-940, 2001.

[8] K. Shung and M. Zippuro, "Ultrasonic transducers and arrays," *IEEE Engineering in Medicine and Biology Magazine,* vol. 15, no. 6, pp. 20-30, 1996.

[9] R. Lu, T. Manzaneque, Y. Yang and S. Gong, "Lithium niobate phononic crystals for tailoring performance of RF laterally vibrating devices," *IEEE Transactions on Ultrasonics, Ferroelectrics, and Frequency Control,* vol. 65, no. 6, pp. 934-944, 2018.

[10] L. Binci, C. Tu, H. Zhu and J.-Y. Lee, "Planar ring-shaped phononic crystal anchoring boundaries for enhancing the quality factor of Lamb mode resonators," *Applied Physics Letters,* vol. 109, no. 20, p. 203501, 2016.

[11] L. Haofeng, J. a. W. L. Rui, C. Chen and L. Xinyu, "Surface acoustic wave sensors of delay lines based on MEMS," *Journal of Nanoscience and Nanotechnology,* vol. 10, no. 11, pp. 7258-7261, 2010.

[12] T. Gorishnyy, C. K. Ullal, M. Maldovan, G. Fytas and E. Thomas, "Hypersonic phononic crystals," *Physical review letters,* vol. 94, no. 11, p. 115501, 2005.

[13] A. Cleland, D. Schmidt and C. S. Yung, "Thermal conductance of nanostructured phononic crystals," *Physical Review B,* vol. 64, no. 17, p. 172301, 2001.

[14] M. Sledzinska, B. Graczykowski, J. Maire, E. Chavez-Angel, C. M. Sotomayor-Torres and F. Alzina, "2D phononic crystals: Progress and prospects in hypersound and thermal transport engineering," *Advanced Functional Materials,* vol. 30, no. 8, p. 1904434, 2020.

[15] M. Merklein, B. Stiller, K. Vu, S. J. Madden and B. J. Eggleton, "A chip-integrated coherent photonic-phononic memory," *Nature Communications,* vol. 8, no. 1, pp. 1-7, 2017.

[16] H. Shin, J. A. Cox, R. Jarecki, A. Starbuck, Z. Wang and P. T. Rakich, "Control of coherent information via on-chip photonic--phononic emitter--receivers," *Nature communications,* vol. 6, no. 1, pp. 1-8, 2015.

[17] E. Verhagen, S. Deleglise, S. Weis, A. Schliesser and T. J. Kippenberg, "Quantum-coherent coupling of a mechanical oscillator to an optical cavity mode," *Nature,* vol. 482, no. 7383, pp. 63-67, 2012.

[18] R. W. Andrews, R. W. Peterson, T. P. Purdy, K. Cicak, R. W. Simmonds, C. A. Regal and K. W. Lehnert, "Bidirectional and efficient conversion between microwave and optical light," *Nature physics,* vol. 10, no. 4, pp. 321-326, 2014.

[19] H. a. Q. W. a. J. R. a. C. J. A. a. O. R. H. a. S. A. a. W. Z. a. R. P. T. Shin, "Tailorable stimulated Brillouin scattering in nanoscale silicon waveguides," *Nature communications,* vol. 4, no. 1, pp. 1-10, 2013.

[20] S. Gertler, P. Kharel, E. A. Kittlaus, N. T. Otterstrom and P. T. Rakich, "Shaping nonlinear optical response using nonlocal forward Brillouin interactions," *New Journal of Physics,* vol. 22, no. 4, p. 043017, 2020.

[21] A. F. Vakakis, O. V. Gendelman, L. A. Bergman, D. M. McFarland, G. Kerschen and Y. S. Lee, Nonlinear targeted energy transfer in mechanical and structural systems, vol. 156, Springer Science \& Business Media, 2008.

[22] Y. Lee, A. F. Vakakis, L. Bergman, D. McFarland, G. Kerschen, F. Nucera, S. Tsakirtzis and P. Panagopoulos, "Passive non-linear targeted energy transfer and its applications to vibration absorption: a review," *Proceedings of the Institution of Mechanical Engineers, Part K: Journal of Multi-body Dynamics,* vol. 222, no. 2, pp. 77-134, 2008.

[23] C. Wang, A. Kanj, A. Mojahed, S. Tawfick and A. F. Vakakis, "Experimental landau-zener tunneling for wave redirection in nonlinear waveguides," *Physical Review Applied,* vol. 14, no. 3, p. 034053, 2020.

[24] C. Wang, A. Kanj, A. Mojahed, S. Tawfick and A. Vakakis, "Wave redirection, localization, and non-reciprocity in a dissipative nonlinear lattice by macroscopic Landau--Zener tunneling: Theoretical results," *Journal of Applied Physics,* vol. 129, no. 9, p. 095105, 2021.

[25] A. Kanj, C. Wang, A. Mojahed, A. Vakakis and S. Tawfick, "Wave redirection, localization, and non-reciprocity in a dissipative nonlinear lattice by macroscopic Landau--Zener tunneling: Experimental results," *AIP Advances,* vol. 11, no. 6, p. 065328, 2021.

[26] M. I. Hussein, M. J. Leamy and M. Ruzzene, "Dynamics of phononic materials and structures: Historical origins, recent progress, and future outlook," *Applied Mechanics Reviews,* vol. 66, no. 4, 2014.

[27] P. D. Garcia, R. Bericat-Vadell, G. Arregui, D. Navarro-Urrios, M. Colombano, F. Alzina and C. M. Sotomayor-Torres, "Optomechanical coupling in the Anderson-localization regime," *Physical Review B,* vol. 95, no. 11, p. 115129, 2017.

[28] J. C. Angel, J. C. T. Guzman and A. D. de Anda, "Anderson localization of flexural waves in disordered elastic beams," *Scientific Reports,* vol. 9, no. 1, pp. 1-10, 2019.

[29] S. Kim, J. Bunyan, P. F. Ferrari, A. Kanj, A. F. Vakakis, A. M. Van Der Zande and S. Tawfick, "Buckling-mediated phase transitions in nano-electromechanical phononic waveguides," *Nano letters,* vol. 21, no. 15, pp. 6416-6424, 2021.

[30] D. Hatanaka, I. Mahboob, K. Onomitsu and H. Yamaguchi, "A phonon transistor in an electromechanical resonator array," *Applied Physics Letters,* vol. 102, no. 21, p. 213102, 2013.

[31] I. Wilson-Rae, R. Barton, S. Verbridge, D. Southworth, B. Ilic, H. G. Craighead and J. Parpia, "High-Q nanomechanics via destructive interference of elastic waves," *Physical review letters,* vol. 106, no. 4, p. 047205, 2011.

[32] J. Liu, K. Usami, A. Naesby, T. Bagci, E. S. Polzik, P. Lodahl and S. Stobbe, "High-Q optomechanical GaAs nanomembranes," *Applied Physics Letters,* vol. 99, no. 24, p. 243102, 2011.

[33] B. Zwickl, W. Shanks, A. Jayich, C. Yang, A. Bleszynski Jayich, J. Thompson and J. Harris, "High quality mechanical and optical properties of commercial silicon nitride membranes," *Applied Physics Letters,* vol. 92, no. 10, p. 103125, 2008.

[34] D. Hatanaka, I. Mahboob, H. Okamoto, K. Onomitsu and H. Yamaguchi, "An electromechanical membrane resonator," *Applied Physics Letters,* vol. 101, no. 6, p. 063102, 2012.

[35] J. Cha and C. Daraio, "Electrical tuning of elastic wave propagation in nanomechanical lattices at MHz frequencies," *Nature nanotechnology,* vol. 13, no. 11, pp. 1016--1020, 2018.

[36] J. Thompson, B. Zwickl, A. Jayich, F. Marquardt, S. Girvin and J. Harris, "Strong dispersive coupling of a high-finesse cavity to a micromechanical membrane," *Nature,* vol. 452, no. 7183, pp. 72--75, 2008.

[37] J. C. Sankey, C. Yang, B. M. Zwickl, A. M. Jayich and J. G. Harris, "Strong and tunable nonlinear optomechanical coupling in a low-loss system," *Nature Physics,* vol. 6, no. 9, pp. 707--712, 2010.

[38] D. Hatanaka, I. Mahboob, K. Onomitsu and H. Yamaguchi, "Phonon waveguides for electromechanical circuits," *Nature nanotechnology,* vol. 9, no. 7, pp. 520-524, 2014.

[39] J. Cha, K. W. Kim and C. Daraio, "Experimental realization of on-chip topological nanoelectromechanical metamaterials," *Nature,* vol. 564, no. 7735, pp. 229--233, 2018.

[40] Z. Bazant and L. Cedolin, "Von Mises Truss," in *Stability of structures: elastic, inelastic, fracture and damage theories*, Singapore, World Scientific, 2010, pp. 228-231.

[41] A. Kanj, P. F. Ferrari, A. M. van der Zande, A. F. Vakakis and S. Tawfick, "Ultra-tuning of nonlinear drumhead MEMS resonators by electro-thermoelastic buckling".

[42] P. W. Anderson, "Absence of diffusion in certain random lattices," *Physical review,* vol. 109, no. 5, p. 1492, 1958.

[43] A. P. Mosk, A. Lagendijk, G. Lerosey and M. Fink, "Controlling waves in space and time for imaging and focusing in complex media," *Nature photonics,* vol. 6, no. 5, pp. 283-292, 2012.

[44] M. Segev, Y. Silberberg and D. N. Christodoulides, "Anderson localization of light," *Nature Photonics,* vol. 7, no. 3, pp. 197-204, 2013.

[45] H. Hu, A. Strybulevych, J. Page, S. E. Skipetrov and B. A. van Tiggelen, "Localization of ultrasound in a three-dimensional elastic network," *Nature Physics,* vol. 4, no. 12, pp. 945-948, 2008.

[46] M. Kurosu, D. Hatanaka, H. Okamoto and H. Yamaguchi, "Buckling-induced quadratic nonlinearity in silicon phonon waveguide structures," *Japanese Journal of Applied Physics,* vol. 61, no. SD, p. SD1025, 2022.

[47] K. C. Balram, M. I. Davanço, J. D. Song and K. Srinivasan, "Coherent coupling between radiofrequency, optical and acoustic waves in piezo-optomechanical circuits," *Nature photonics,* vol. 10, no. 5, pp. 346-352, 2016.

[48] K.-T. Wan, S. Guo and D. A. Dillard, "A theoretical and numerical study of a thin clamped circular film under an external load in the presence of a tensile residual stress," *Thin Solid Films,* vol. 425, no. 1-2, pp. 150--162, 2003.

[49] G. W. Vogl and A. H. Nayfeh, "A reduced-order model for electrically actuated clamped circular plates," *International Design Engineering Technical Conferences and Computers and Information in Engineering Conference,* vol. 37033, pp. 1867--1874, 2003.